\documentclass[letterpaper,11pt,twoside,twocolumn]{IEEEtran}

\usepackage{times}

\usepackage{graphicx,subfigure}
\usepackage[english]{babel}
\usepackage{amssymb}
\usepackage{url}                      
\usepackage[super,negative]{nth}      
                                     
\usepackage{icomma}

\usepackage[ruled]{algorithm2e}

\newcommand{\co}[2]{#1 \tiny{$\pm#2$}}

\begin{document}

\title{DTN Routing in a Mobility Pattern Space}

\author{\large J\'er\'emie Leguay$^{*+}$, Timur Friedman$^{*}$, Vania Conan$^{+}$\\
  \\
  \normalsize $^{*}$Universit\'e Pierre et Marie Curie, Laboratoire LiP6--CNRS\\
  \normalsize $^{+}$Thales Communications \thanks{This work has been
    co-funded by Thales Communications, LIP6 and the ANRT (Association
    Nationale de la Recherche Technique) through the CIFRE grant
    135/2004.}  }

\maketitle

\begin{abstract}
  Routing in Delay Tolerant Networks (DTNs) benefits considerably if
  one can take advantage of knowledge concerning node mobility. The main
  contribution of this paper is the definition of a generic routing
  scheme for DTNs using a high-dimensional Euclidean space constructed
  upon nodes' mobility patterns. For example, nodes are represented
  as points having as coordinates their probability of being found in
  each possible location. We present simulation results indicating
  that such a scheme can be beneficial in a scenario inspired by
  studies done on real mobility traces.  This work should open the way
  to further use of the virtual space formalism in DTN routing.
\end{abstract}

\section{Introduction}\label{sec_introduction}

The novelty of the work presented here is that we re-transcribe the
problem of routing in a Delay Tolerant Network
(DTN)~\cite{dtn_fall_sigcomm} on the basis of mobility patterns into a
problem of routing in a virtual space defined by those patterns.  By
doing so, we can bring the powerful formalism of Euclidean space to
bear on the problem of DTN routing.

In one common DTN scenario, nodes are mobile and have wireless
networking capabilities.  They are able to communicate together only
when they are within transmission range. The network suffers from
frequent connectivity disruptions, making the topology intermittently
and partially connected. This means that there is a very low
probability that an end-to-end path exists between a given pair of
nodes at a given time.  End-to-end paths can exist temporarily, or may
sometimes never exist, with only partial paths emerging.  Due to these
disruptions, regular ad-hoc networking approaches to routing and
transport do not hold, and new solutions must be proposed.

The Delay Tolerant Network Research Group (DTNRG)~\cite{dtnrg} has
proposed an architecture~\cite{draftirtf} to support messaging that
may be used by delay tolerant applications in such a context. The
architecture consists mainly of the addition of a new layer to the
regular TCP/IP stack called the bundle layer. Messages transferred in
DTNs are called bundles. They are transferred in an atomic fashion
between nodes using TCP in order to ensure node-to-node reliability.
These messages can be of any size. Nodes are assumed to have large
buffers in which they can store the bundles.

Routing is one of the very challenging open issues in DTNs, as
mentioned by Jain et al.~\cite{routingdtn}. Indeed, since the network
suffers from connectivity issues, MANET~\cite{rfc2501} routing
algorithms such as OLSR, based on the spreading of control
information, or AODV, which is on-demand, fail to achieve routing.
Different approaches have to be found.

Epidemic routing is a possible solution when nothing is known about
the behavior of nodes. Since it can lead to buffer overloads and
inefficient use of transmission media, one would prefer to limit
bundle duplication and instead use routing heuristics that can take
advantage of context. In order to move in such a direction, the DTN
architecture defines several types of contacts: scheduled,
opportunistic, and predicted. \textit{Scheduled} contacts can exist,
for instance, between a base station somewhere on earth and a low
earth orbiting relay satellite.  \textit{Opportunistic} contacts are
created simply by the presence of two entities at the same place, in a
meeting that was neither scheduled nor predicted. Finally,
\textit{predicted} contacts are also not scheduled, but predictions of
their existence can be made by analyzing previous observations.

Some work has been done with scheduled contacts, such as the paper by
Jain et al.~\cite{routingdtn} that tries to improve the connectivity
of an isolated village to the internet based on knowledge of when a
low-earth orbiting relay satellite and a motor bike might be available
to make the necessary connections. Also of interest, work on
inter-planetary~\cite{ipn} networking uses predicted contacts such as
the ones between planets within the framework of a DTN architecture.
The case of only opportunistic contacts has been analyzed by Vahdat
and Becker~\cite{vahdat00epidemic} using the epidemic routing scheme.
Most of the work around routing in DTNs has been performed with
predicted contacts, such as the algorithm of Lindgren et
al.~\cite{lindgren03}, which relies on nodes having a community
mobility pattern.  Nodes mainly remain inside their community and
sometimes visit the others.  As a consequence, a node may transfer a
bundle to a node that belongs to the same community as the
destination. In a similar manner, Burns et al.~\cite{mvrouting}
proposed a routing algorithm that uses past frequencies of contacts.
Also making use of past contacts, Davis et al.~\cite{davis01} improved
the basic epidemic scheme with the introduction of adaptive dropping
policies. Recently, Musolesi et al.~\cite{musolesi05} have introduced
a generic method that uses Kalman filters to combine and evaluate the
multiple dimensions of the context in order to take routing decisions.
The context is made of measurements that nodes perform periodically,
which can be related to connectivity issues, but not necessarily.
This mechanism allows network architects to define their own hierarchy
among the different context attributes.

The case study presented in this paper relies also on contacts that
can be characterized as predicted, but the underlying idea is a more
generic abstraction compared to previous work, being able to capture
the interesting properties of major mobility patterns for routing.
The main contribution of this paper is the use, for routing in DTNs,
of the formalism of a high-dimensional Euclidean space based on nodes'
mobility patterns.  We show the feasibility of this concept through an
example in which each dimension represents the probability for a node
to be found in a particular location. We conduct a simulation that
produces promising initial results for this concept.

Applying the formalism of Euclidean space to computer networking
problems is not in itself a new idea.  To our knowledge, however, it
has not previously been used for DTN routing.  Furthermore, we believe
that the idea of constructing a virtual space based upon mobility
patterns is new.

Previous work with Euclidean spaces for networking has included the
geolocalization of internet hosts, as in the GeoPing technique
developed by Padmanabhan and
Subramanian~\cite{padmanabhan01investigation}. The position of a host
to localize in Euclidean space is compared to the position of well
know landmark nodes in order to estimate the host's location.  Round
trip times are used to determine coordinates. As opposed to what we do
here, this Euclidean space is not used for routing.

Euclidean spaces have also been exploited in peer-to-peer
architectures, notably by Ratnasamy et al.\ for
CAN~\cite{ratnasamy00scalable}, in order to construct a robust and
scalable mechanism to handle search queries.  In this case, the
Euclidean space is a virtual space, in which keys describing files are
assigned virtual coordinates, and each node in the system governs a
portion of the space.  Queries are routed from node to node in the
direction of the key.

The routing scheme presented in this paper is similar to the one of
CAN, since messages are routed in a virtual space. As opposed to CAN,
in our scheme there is no notion of neighbor in the virtual space.
Nodes may be directly connected to nodes that are nearby or that are
very far. And these connections arise and dissolve dynamically as a
function of node mobility in the physical space. Nodes
opportunistically take advantage of connections that promise to
advance bundles toward the destination.

The rest of the paper is structured as follows.  Sec.~\ref{sec_algo}
describes the mobility pattern based routing scheme.
Sec.~\ref{sec_simus} presents the simulation results.
Sec.~\ref{sec_conclu} concludes the paper, discussing directions for
future work.

\section{Routing in a mobility pattern space}\label{sec_algo}

This section first presents the idea behind routing in a
high-dimensional Euclidean space constructed upon mobility patterns of
nodes and then shows how we applied this idea within the framework of
a scenario inspired by real observations.

\subsection{Concept}\label{concept}

The Euclidean virtual space introduced here is a generalization of
ideas that are already current in the DTN literature.  The principle
is to use a Euclidean space as a tool to help nodes to take routing
decisions.  These decisions rely on the notion that a node is a good
candidate for taking custody of a bundle if it has a mobility pattern
similar to that of the bundle's destination.
Routing is done by forwarding bundles toward nodes that have mobility
patterns that are more and more similar to the mobility pattern of the
destination.

In the space that we define, the mobility pattern of a node provides
its coordinates.
Several questions arise. 
What type of dimensions do we choose, how many, and what kind of range
for values do we define? How do we define the notion of distance? Is
straightforward Euclidean distance useful or are other similarity
functions more appropriate?  Is it possible to have an infinite space
in terms of the number of dimensions?  What might be the problems with
such a scheme?

Note that the objective of this paper is not to answer all these
questions. It is to introduce a new concept for routing and to examine
some of the interesting problems that the concept presents.  In this
section, we describe a manner in which mobility patterns can be
characterized and discuss other possible alternatives.  The question
of the choice of a similarity metric is addressed in
Sec.~\ref{case_study}.

\subsubsection{Mobility pattern characterization}

Several methods could be employed to describe a mobility pattern. For
instance it could be based upon historic information regarding
contacts that the node has already had. 
If we want to route a bundle from one node to another, we have an
interest in considering information on these contacts as forming a
virtual space. Each possible contact is an axis, and the distance
along that axis indicates a measure of the probability of contact.
Two nodes that have a similar set of contacts that they see with
similar frequencies are close in this space, whereas nodes that have
very different sets of contacts, or that see the same contacts but
with very different frequencies, are going to be far. It seems
reasonable that one would wish to pass a bundle to a node that is as
close as possible to the destination in this space, because this
should improve the probability that it will eventually reach the
destination.

In the virtual contact space just described, knowledge of all axes of
the space requires knowledge of all nodes that are circulating in the
space.  Note that a full knowledge of the axes might not be required
for successful routing.  Nonetheless, we might wish to consider an
alternative space in which there is a fixed, or at least more limited
and well known number of axes.  If nodes' visits to particular
locations can be tracked, then the mobility pattern of a node can be
described by its visits to these locations.  In this scenario, each
axis represents a location, and the distance along the axis represents
the probability of finding a node at that location.  We can imagine
that nodes that have similar probabilities of visiting a similar set of
locations are more likely to encounter each other than nodes that are
very different in these respects.  This is the virtual space that we
employ in the study described in Sec.~\ref{case_study}.

\subsubsection{Possible limits and issues}

DTN Routing in a contact space or a mobility space is based on the
assumption that there will be regularities in the contacts that nodes
have or their choices of locations to visit.  There is always the
possibility that, we may encounter mobility patterns similar to the
ones observed with random mobility models.  The efficiency of the
virtual space as a tool may be limited if nodes too rapidly change
their habits.

Some problems could occur even if nodes have well defined mobility
patterns 
For instance, in the Euclidean space, a bundle may reach a local
maximum if a node has a mobility pattern that is the most similar in
the local neighborhood to the destination node's mobility pattern, but
is not sufficient for one reason or another to achieve the delivery.
In the second type of space where each dimension represents a
location, it can happen if nodes visit similar places, but for timing
reasons, such as being on opposite diurnal cycles, they never meet.

Other issues surrounding the use of a virtual contact space or
mobility space are discussed in Sec.~\ref{sec_conclu}.

\subsection{A case study}\label{case_study}

Recent studies of the mobility of students in a campus
\cite{ucsd,dartmount,intel} or corporate users
\cite{balazinska2003characterizing} equipped with PDAs or laptops able
to be connected to wireless access networks, show that they follow
common mobility patterns.  They show that significant aspects of the
behavior can be characterized by power-law distributions.
Specifically, the session durations and the frequencies of the places
visited by users follow power laws.  This means that users typically
visit a few access points frequently while visiting the others rarely,
and that users may stay at few locations for long periods while
visiting the others for very short periods. Henderson et al.
observed~\cite{dartmount} that $50$\% of users studied spent $62$\% of
their time attached to a single access point and this proportion
decreased exponentially.  If we take these wireless access network
studies to be representative of a class of mobile node behavior, we
can consider that these observations are applicable to at least
certain DTN scenarios.

For this case study we propose the following mobility model. Let us
consider a set of nodes that move among a set of $N$ locations. Two
nodes can communicate only if they are at the same location.  Node
movements are based on power-laws, and each node has a mobility
pattern defined by the distribution of $P$. $P(i)$ is the probability
for the node to be at location $i$ and $P(i) = K \left( \frac{1}{d}
\right)^{n_i}$ with $n_i$ the probability of being found in the
location $i$, $d$ the exponent of the power-law based mobility pattern
and $K$ a constant. $n_j=0$ means that the location $j$ is the
preferred one. Because $\sum_ {i} P(i)=1$, we have:

\begin{equation}
K=\frac{1-\frac{1}{d}}{1-\frac{1}{d^{N}}}
\label{eq_d} 
\end{equation}

\begin{figure}[!h]
\begin{center}
\includegraphics[width=8cm]{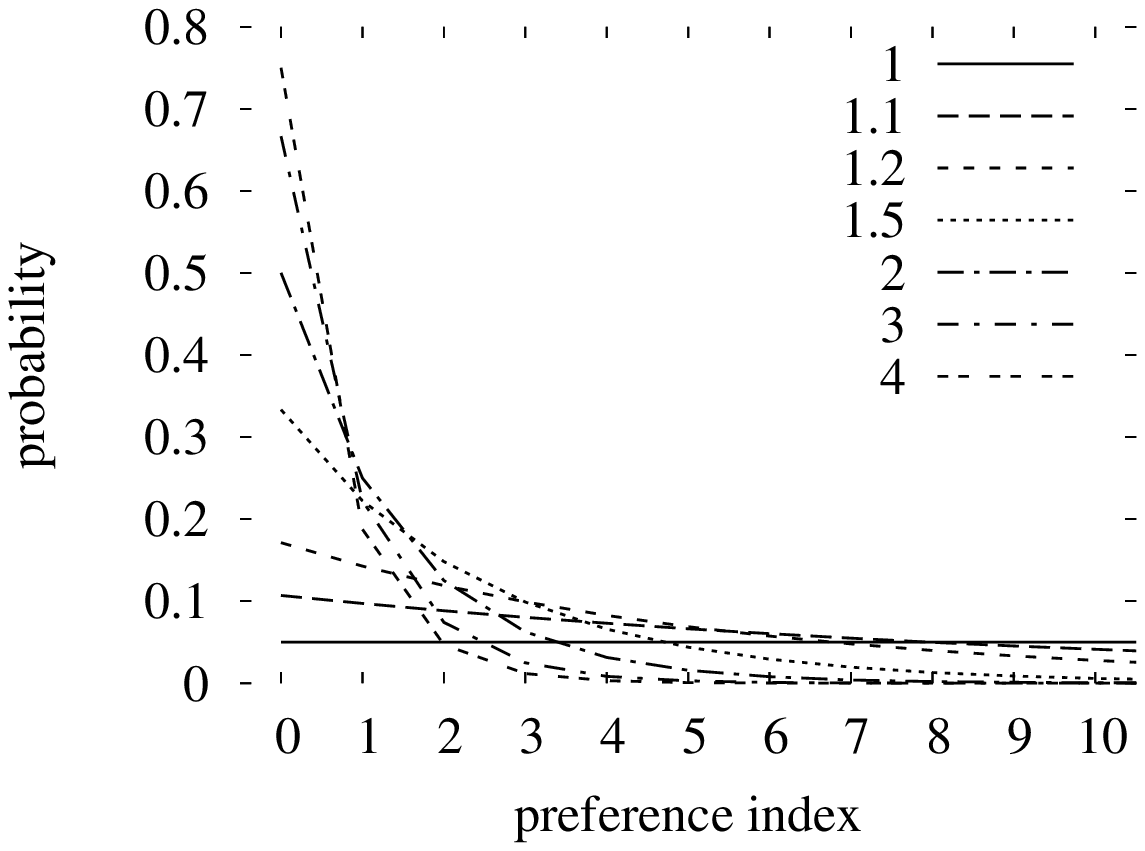}
\caption{\label{im_distributions} Power-law probability distributions for different values of $d$.}
\end{center}
\end{figure}

Under this model, $d$ is the fundamental parameter governing node
behavior.  As shown in Fig.~\ref{im_distributions}, when $d$ is high,
nodes tend to move among a very small subset of locations, having one
that they strongly prefer to the others. As $d$ approaches $1$, the
range of locations that nodes visit regularly becomes wider, while
still presenting a hierarchy of preferences.  When $d=1$, we have
equiprobability.

Note that a mobility pattern should also characterize the time that a
node remains at each location. We propose to capture this in the
mobility model described here by uniformly distributing the resting
time at each location over the interval
$[t_\mathrm{min},t_\mathrm{max}]$ with $t_\mathrm{min}$ and
$t_\mathrm{max}$ quite close together, and by allowing nodes to
randomly choose the same location consecutively.

For each of the nodes in this model, there is therefore a well defined
probability of finding that node at each of the $N$ locations.  This
set of probabilities is a node's mobility pattern, and is described by
a point in an $N$ dimensional Euclidean space.  We propose to route
bundles in this space by sending them to nodes having mobility
patterns that are successively closer to the mobility pattern of the
destination. In simple words, we prefer to give custody of a bundle to
a node that has mobility habits similar to those of the bundle's
destination. In order to complete our model, we therefore require a
similarity function that can be used to compare mobility patterns.

We study the following functions, each of which is a measure of
similarity between two points in a Euclidean space:

\begin{itemize}
  
\item \textit{Euclidean distance}: This is the most common distance
  measure. It returns the root of the sum of the square differences
  between the coordinates of a pair of points.

\begin{equation}
d_{ij}=\sqrt{\sum_ {k=1}^{n} \left(x_{ik}-x_{jk} \right)^2}
\end{equation}

\item \textit{Canberra distance}: Canberra distance returns the sum of
  a series of fractional differences between coordinates of a pair of
  points. Each fractional difference term has a value between 0 and 1.
  If one of coordinate is zero, the term is defined to be unity
  regardless the other value.

\begin{equation}
d_{ij}=\sum_ {k=1}^{n} \frac{\left| x_{ik}-x_{jk} \right|} { \left|x_{ik}\right|+\left|x_{jk} \right| }
\end{equation}

\item \textit{Cosine angle separation}: This measure represents the
  cosine of the angle between two vectors. It measures similarity
  rather than distance or dissimilarity. Thus, a high cosine angle
  separation value indicates that two vectors are similar.

\begin{equation}
s_{ij}=\frac{\sum_ {k=1}^{n} x_{ik}.x_{jk} }{\sqrt{\sum_ {k=1}^{n} x_{ik}^2 . \sum_ {r=1}^{n} x_{jr}^2 }}
\end{equation}

\item \textit{Matching distance}: This measure is simply the raw
  number of location probabilities that are similar for two nodes.  We
  consider two coordinates on a given axis to be similar if their
  absolute difference is less than or equal to a defined value $\delta$.\\

\end{itemize}

In the routing scheme presented here, the only information that must
be flooded by nodes is their mobility patterns. These mobility
patterns can be spread in an epidemic fashion.
Furthermore, we introduce an optimization based upon the power-law
distribution of probabilities.  Under this optimization, nodes
transmit only the main components of their mobility patterns. The
other components are presumably negligible in comparison.  We examine
the performance and the limits of such an optimization in
Sec.~\ref{partial}.

\section{Simulation results}\label{sec_simus}

This section presents the manner in which we evaluated the mobility
pattern based routing scheme, and the results we obtained.

\subsection{Methodology}\label{sec_method}

We have implemented a stand alone simulator to evaluate the mobility
pattern based routing scheme presented in this paper. This simulator
only implements the transport and network layers and it makes simple
assumptions regarding lower layers, for instance allowing for infinite
bandwidth.

In this paper, we refer to routing based on mobility patterns by the
following names: \emph{Euclidean}, using the Euclidean distance
metric; \emph{Canberra}, using the Canberra distance metric;
\emph{Angle}, using the cosine angle separation distance metric; and
\emph{Matching}, using the matching distance metric.  We compared
these routing algorithms against the following:

\begin{itemize}
  
\item \textit{Epidemic}: This is based on epidemic routing, as
  described by Vahdat and Becker~\cite{vahdat00epidemic}: Each time
  two nodes meet, they exchange their bundles. The major interest of 
  this algorithm is that it
  provides the optimum path and thus the minimum bundle delay.  We use
  it here as a lower bound.  In practice, epidemic routing suffers
  from high buffer occupancy and high bandwidth utilization.
  
\item \textit{Opportunistic}: A node waits to meet the destination in
  order to transfer its bundle. The main advantage of this method is
  that it involves only one transmission per bundle. 
  
\item \textit{Random}: When a node is at a location and the bundle's
  destination in not there, the node transfers the bundle to a
  neighbor chosen at random.  We have added a rule to avoid local
  loops: a node can only handle a bundle one time per location visit.
  This scheme is used in this paper as another basis of comparison.  A
  novel algorithm should perform better than this one in order to be
  valuable.

\end{itemize}

All the scenarios simulated in the rest of the paper share common
parameters that can be found in Table~\ref{simulation_parameters}.  We
considered a set of $25$ locations. The virtual space used for routing
thus has $25$ dimensions.  There are $50$ mobile nodes.  Every node
generates bundles destined toward each of the others every $30$s with
the first bundle being sent at a time randomly chosen from a uniform
distribution over the interval $[0,30\mathrm{s}]$.  Simulations last $4000$s.
We generate traffic in the first $500$s of the simulations in
order to give enough time for all the bundles to reach their
destination. The simulator used a time step of $10$ms.

\begin{table}[!h]
\begin{center}
\scriptsize
\begin{tabular}{|l|c|}
\hline
\textbf{Parameter} & \textbf{Value}\\
\hline
Number of nodes & 50 \\
Number of locations & 25 \\
Simulation duration & 4000s \\
Traffic generation & until 500s \\
Packet interval & 30s \\
$t_\mathrm{min}$ & 5s \\
$t_\mathrm{max}$ & 15s \\
$\delta$ & $2 10^{-8}$  \\
Time step & 10 ms \\
\hline
\end{tabular}
\end{center}
\caption{\label{simulation_parameters}Simulation parameters.}
\end{table}

We have tested two variants of the mobility pattern based routing
scheme.  In the first, we assume that a node that is sending a bundle
has full knowledge of the destination's mobility pattern, and that it
addresses the bundle accordingly.  In the second, we assume that nodes
communicate only the major components of their mobility patterns.
This reduces the amount of control traffic exchanged between nodes,
but it also means that a node that is sending a bundle can only
specify partial information regarding the destination.

\subsection{Results}\label{sec_res}

We evaluate routing algorithms on their transport layer performance in
the simulation. We consider a good algorithm to be one that yields a
low average bundle delay and a low average route length.\\

\subsubsection{With full knowledge}\label{sec_full}

We preface our detailed discussion of simulation results with the
observation that Euclidean and Angle yielded identical results.  This
may be explained by the fact that when the number of dimensions of the
space is high, there is a strong correlation between those metrics, as
shown by Qian et al.~\cite{similarity}. In this section, therefore,
the Euclidean and Angle results are reported together.

We performed $5$ runs for each set of parameters (the number was
limited by the length of time required for simulations). Figures
reported in the tables here are mean results with confidence intervals
at the $90$\% confidence level, obtained using the Student $t$
distribution.

\begin{table}[!h]
\begin{center}
\scriptsize
\begin{tabular}{|c|c|c|c|}
\hline
$d$  & 1.1 & 1.5 & 2 \\
\hline
\textbf{Epidemic} & \co{10.9}{7.3}  & \co{13.2}{0.4} & \co{16.2}{0.5}  \\
\textbf{Opportunistic} & \co{123.3}{7.7} & \co{287.4}{8.4} & \co{550.2}{15.2}  \\
\textbf{Random} & \co{117.8}{8.0} & \co{160.0}{1.9} & \co{203.3}{17.3} \\
\textbf{Euclidean \& Angle} & \co{103.0}{7.7} & \co{59.1}{2.7} & \co{54.6}{2.0} \\
\textbf{Canberra} & \co{104.8}{4.6} & \co{113.4}{10.4} & \co{245.0}{41.2}  \\
\textbf{Matching} & \co{118.5}{5.7} & \co{189.5}{12.1} & \co{352.9}{56.0} \\
\hline
\end{tabular}
\end{center}
\caption{\label{res_avgdelay}Average bundle delay.}
\end{table}

Table~\ref{res_avgdelay} presents the mean bundle delay obtained for
each routing algorithm, and for various exponents, $d$, of the power
law distribution, describing the preferential attachment of nodes
toward each location. The notable feature of these results is that
Euclidean and Angle show improved performance with an increase in $d$,
whereas performance declines for all other routing algorithms as $d$
increases.  Opportunistic performs worst, followed closely by Random,
Matching, and Canberra. The fact that Matching and Canberra are worse
than Random is interesting. One hypothesis could be that they are
actively making poor choices.  However, we have reason to believe that
Random has a delay advantage that Matching and Canberra do not share.
In Random, bundles will jump to other nodes without any preference
ordering.  This makes for highly mobile bundles, as is borne out by
their extraordinarily high average route lengths, shown in
Table~\ref{res_avglength}.  One might not necessarily want to pay the
price of such processing overhead in order to obtain modest gains in
delay.  A better standard for comparison might be a random algorithm
that shows preferences, as do Matching and Canberra, but preferences
that are purely random in nature.  Our judgment concerning Matching
and Canberra is thus still in reserve pending further study.

Projecting from the results in this section, we might ask what would
happen for ever higher values of $d$.  Recall that, the higher the
value of $d$, the higher the probabilities are that nodes will find
themselves at a few select locations.  For a high enough value, there
would be little node movement, and little diversity in their
movements.  We would expect this to have a negative effect on all
routing schemes, including Euclidean and Angle.  We have not yet had
the opportunity to conduct studies to see if this phenomenon emerges
as expected.

\begin{table}[!h]
\begin{center}
\footnotesize
\begin{tabular}{|c|c|c|c|}
\hline
$d$  & 1.1 & 1.5 & 2 \\
\hline
\textbf{Epidemic} & \co{3,7}{0.0} & \co{3.7}{0.0} & \co{3.8}{0.1} \\
\textbf{Opportunistic} & \co{1}{0.0} & \co{1}{0.0} & \co{1}{0.0} \\
\textbf{Random} & \co{44.5}{0.7} & \co{55.9}{1.0} & \co{69.8}{2.2} \\
\textbf{Euclidean \& Angle} & \co{3.3}{0.0} & \co{3.2}{0.0} & \co{3.2}{0.0} \\
\textbf{Canberra} & \co{3.3}{0.0} & \co{3.2}{0.0} & \co{3.2}{0.0} \\
\textbf{Matching} & \co{2.5}{0.0} & \co{2.5}{0.0} & \co{2.4}{0.0} \\
\hline
\end{tabular}
\end{center}
\caption{\label{res_avglength}Average route lengths.}
\end{table}

\begin{table}[!htbp]
\begin{center}
\footnotesize
\begin{minipage}[t]{8cm}
\hspace{1cm} \textbf{$d = 1.1 $} \hspace{1.7cm} \textbf{$d = 1.5 $} \hspace{1.8cm} \textbf{$d = 2 $} \hspace{2cm}\\
\vspace{-0.8cm}
\center{\rule{8cm}{0.01cm}}
\vspace{0.1cm}
\end{minipage}
\begin{minipage}[t]{8cm}
\includegraphics[width=2.6cm]{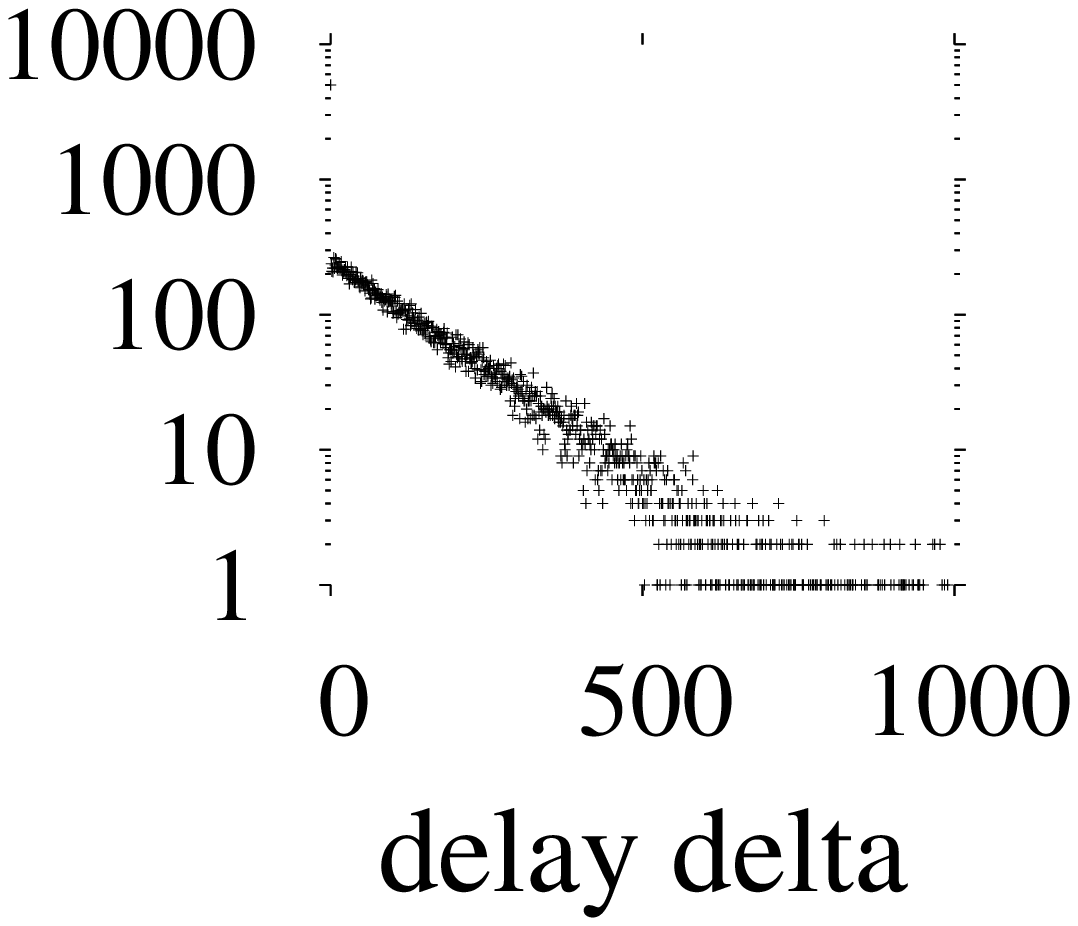}
\includegraphics[width=2.6cm]{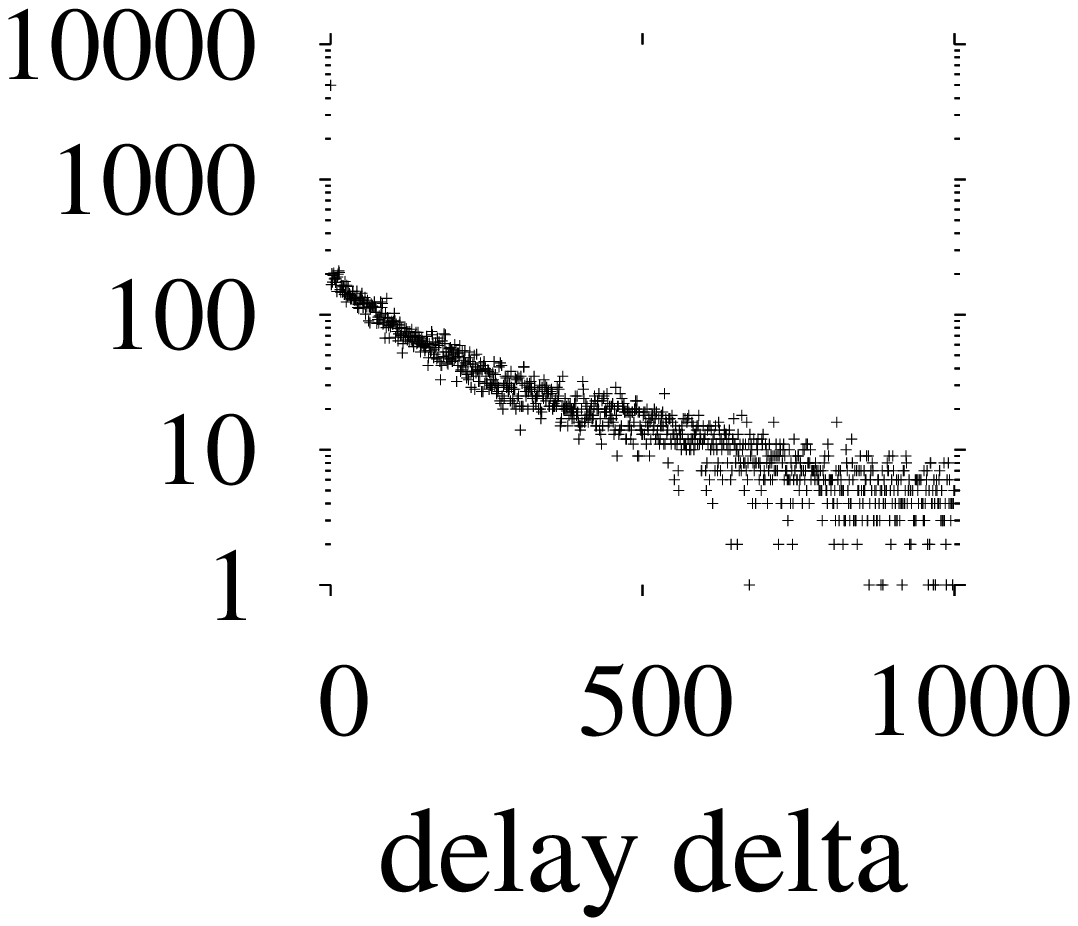}
\includegraphics[width=2.6cm]{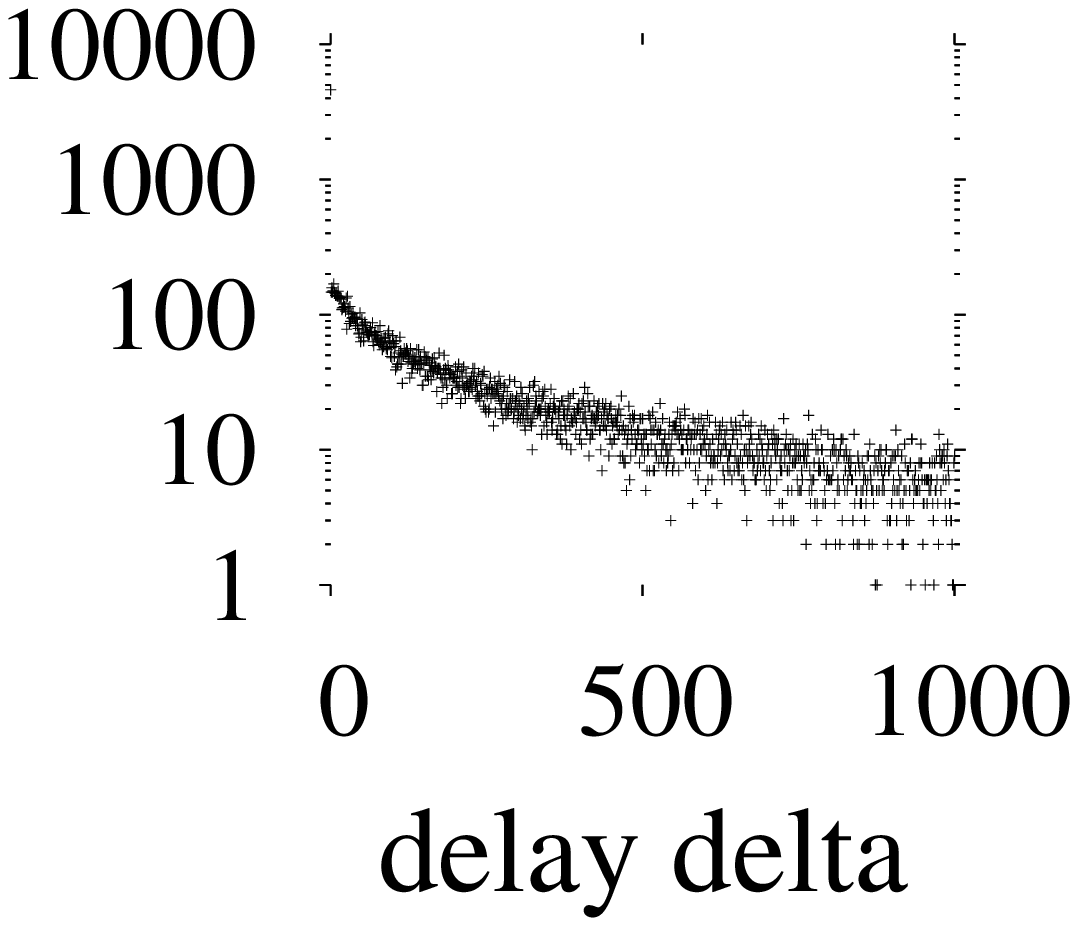}
\end{minipage}
\begin{minipage}[t]{8cm}
\center{\textbf{Opportunistic}\\
\vspace{-0.3cm}
\rule{8cm}{0.01cm}}
\vspace{0.1cm}
\end{minipage}
\begin{minipage}[t]{8cm}
\includegraphics[width=2.6cm]{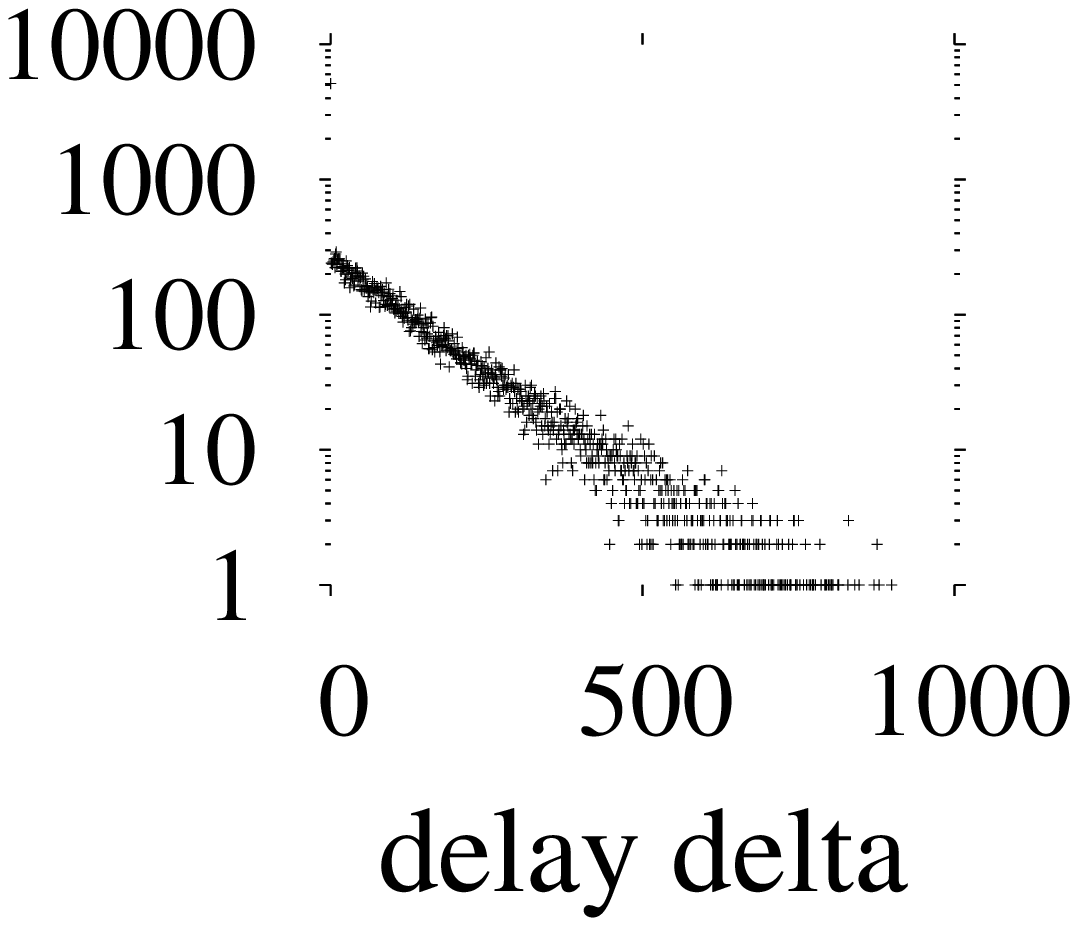}
\includegraphics[width=2.6cm]{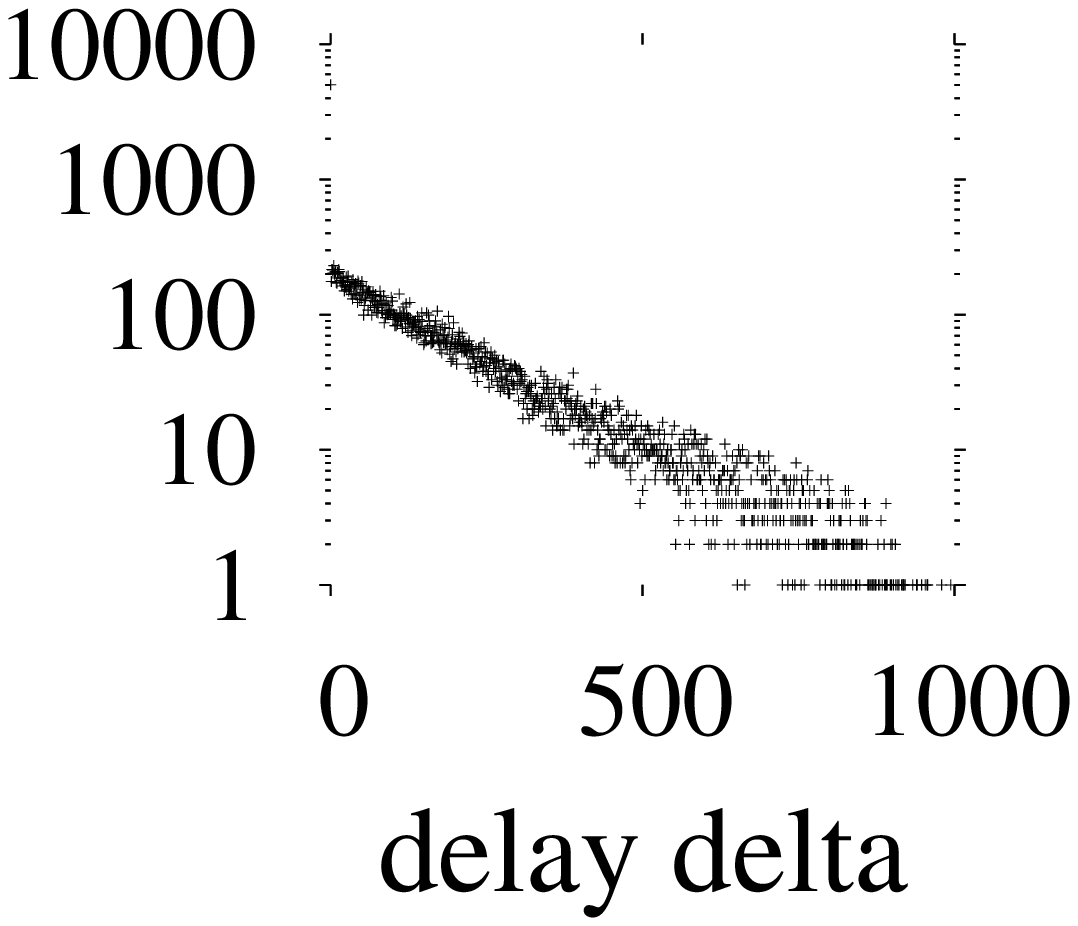}
\includegraphics[width=2.6cm]{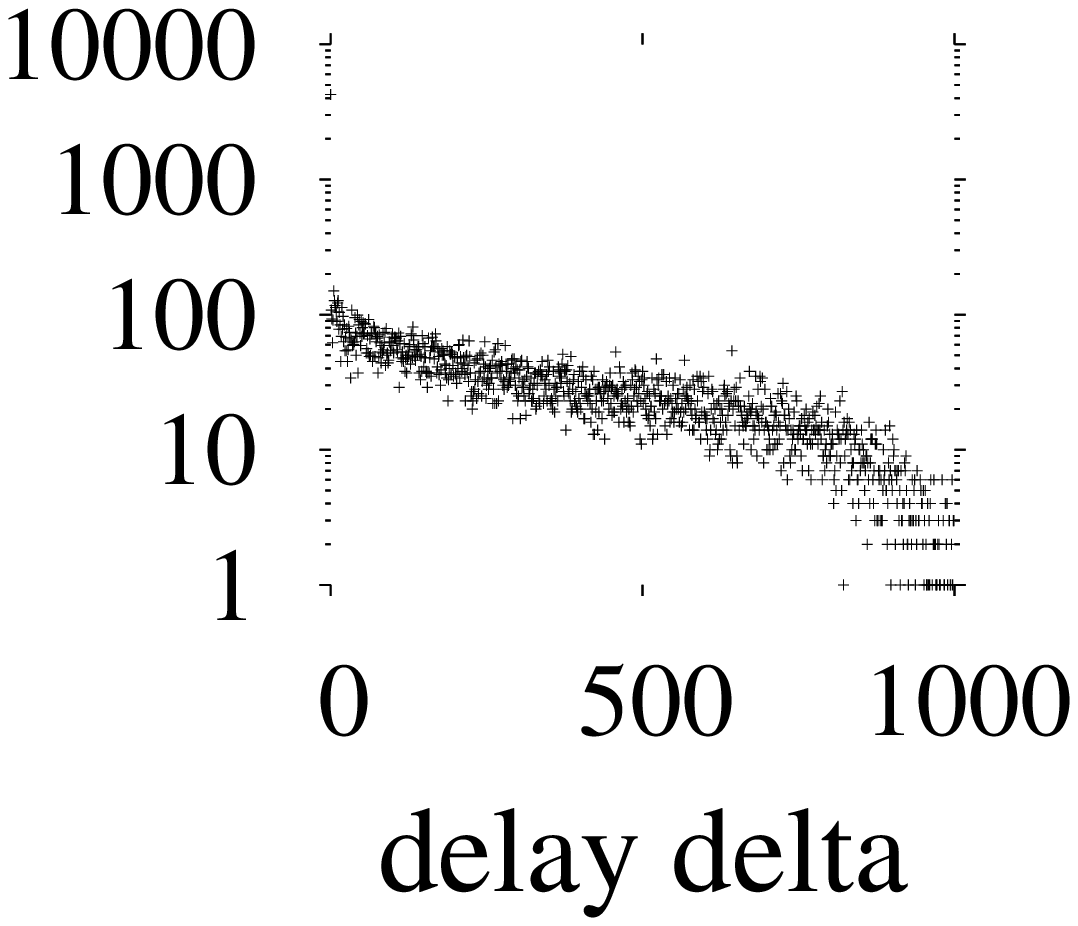}
\end{minipage}
\begin{minipage}[t]{8cm}
\vspace{-0.3cm}
\center{\textbf{Random}\\
\vspace{-0.3cm}
\rule{8cm}{0.01cm}}
\vspace{0.1cm}
\end{minipage}
\begin{minipage}[t]{8cm}
\includegraphics[width=2.6cm]{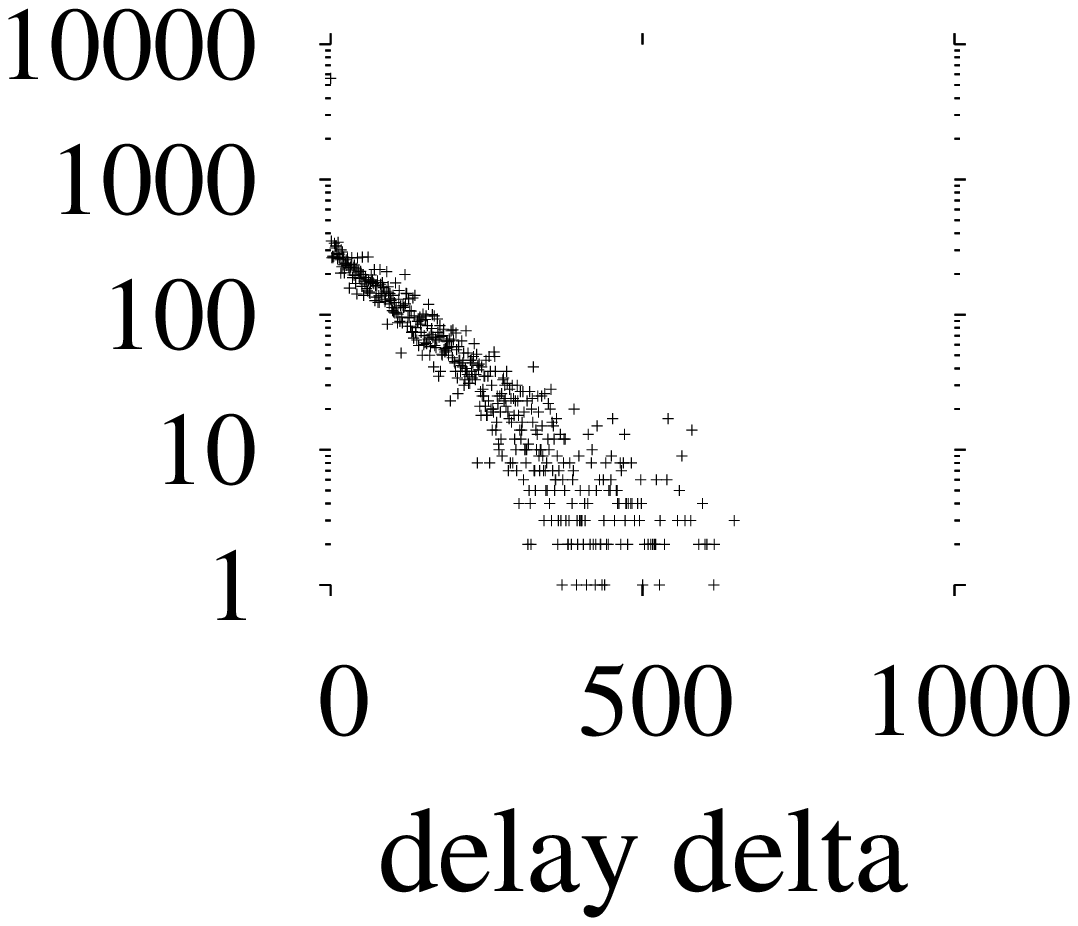}
\includegraphics[width=2.6cm]{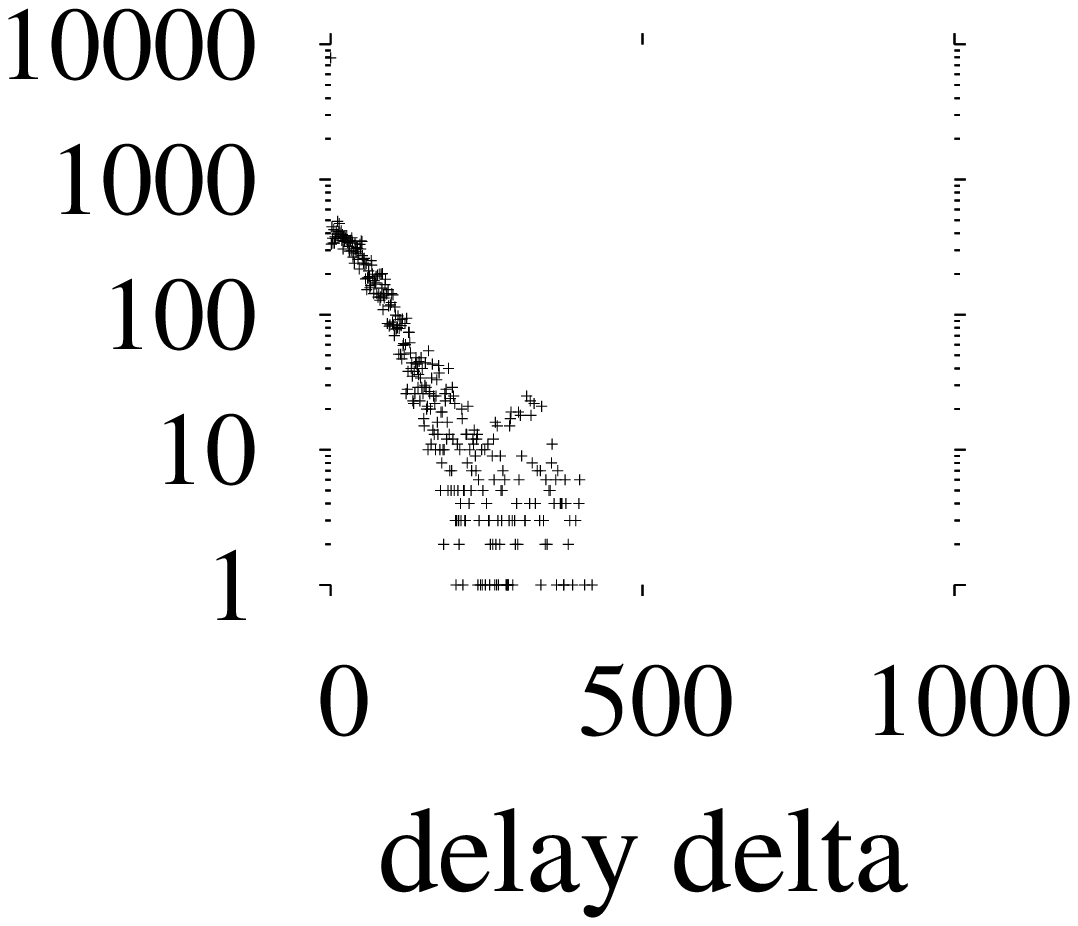}
\includegraphics[width=2.6cm]{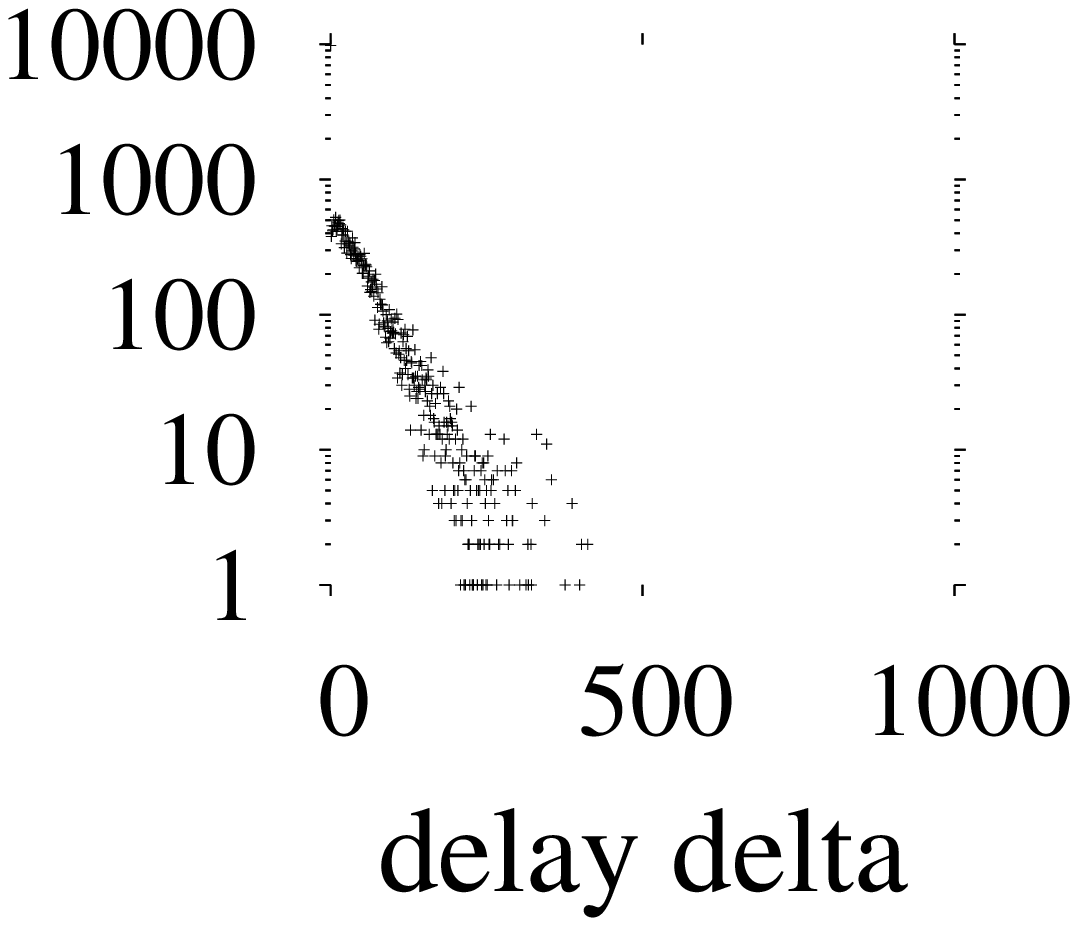}
\end{minipage}
\begin{minipage}[t]{8cm}
\vspace{-0.3cm}
\center{\textbf{Euclidean \& Angle}\\
\vspace{-0.3cm}
\rule{8cm}{0.01cm}}
\vspace{0.1cm}
\end{minipage}
\begin{minipage}[t]{8cm}
\includegraphics[width=2.6cm]{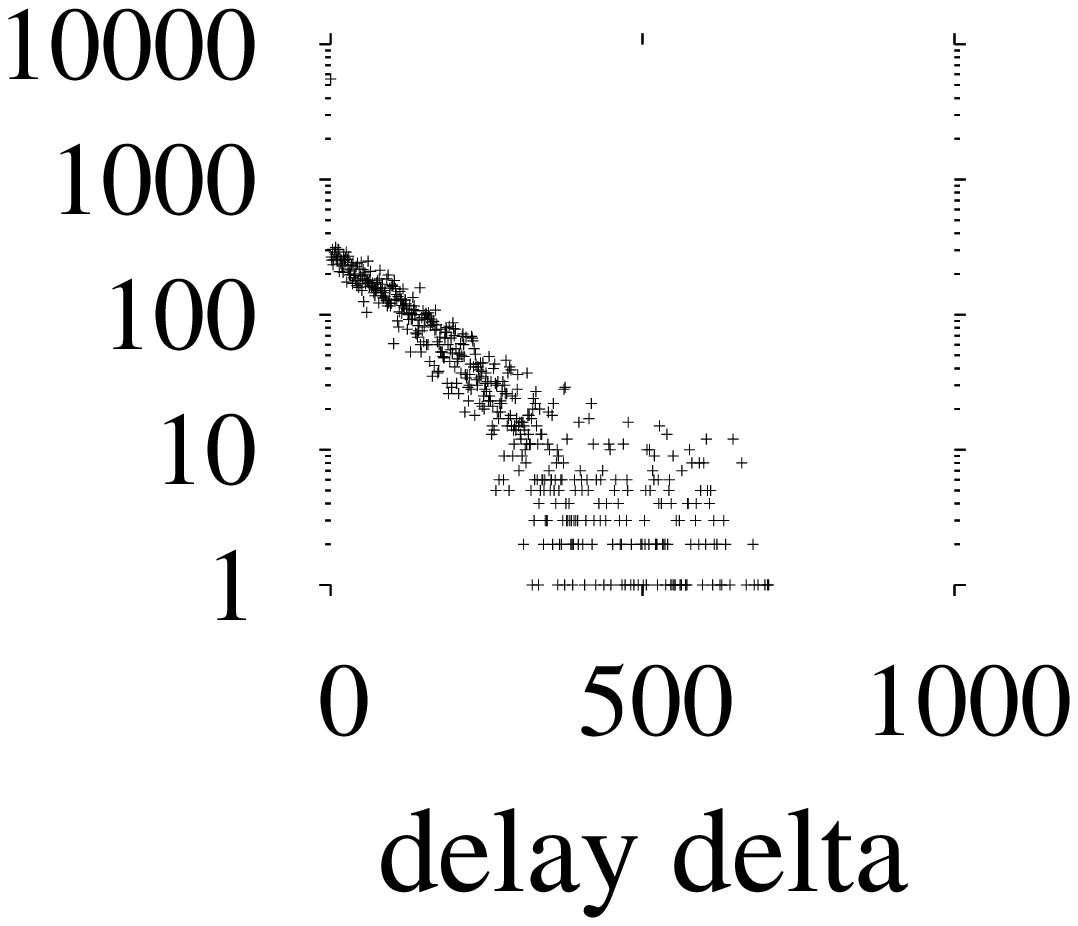}
\includegraphics[width=2.6cm]{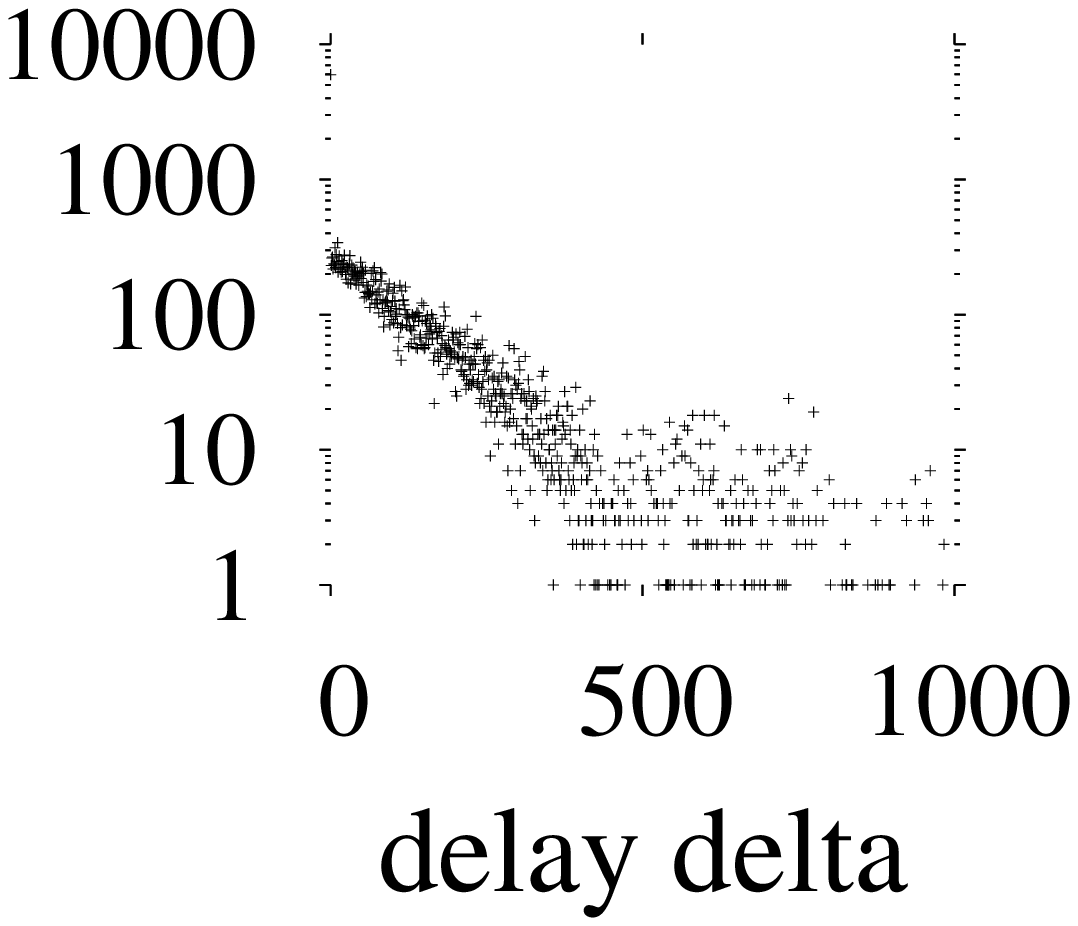}
\includegraphics[width=2.6cm]{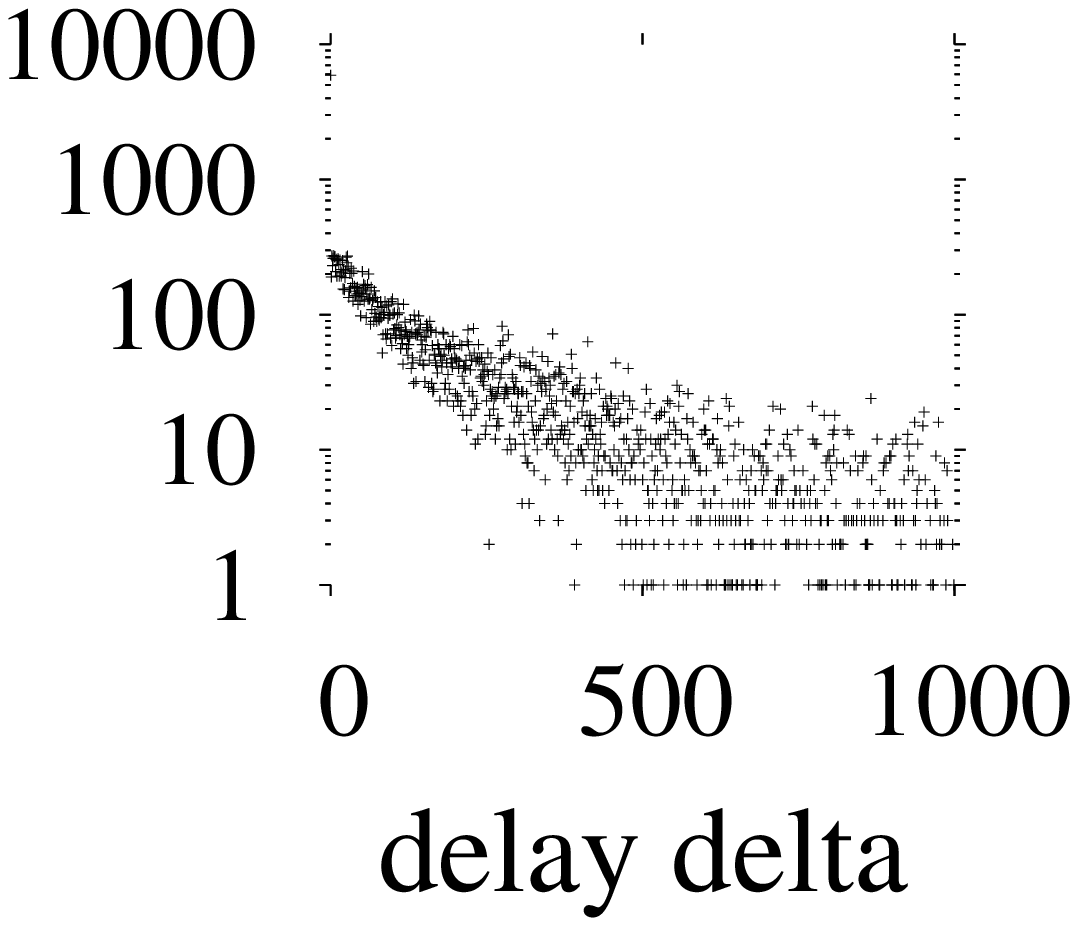}
\end{minipage}
\begin{minipage}[t]{8cm}
\vspace{-0.3cm}
\center{\textbf{Canberra}\\
\vspace{-0.3cm}
\rule{8cm}{0.01cm}}
\vspace{0.1cm}
\end{minipage}
\begin{minipage}[t]{8cm}
\includegraphics[width=2.6cm]{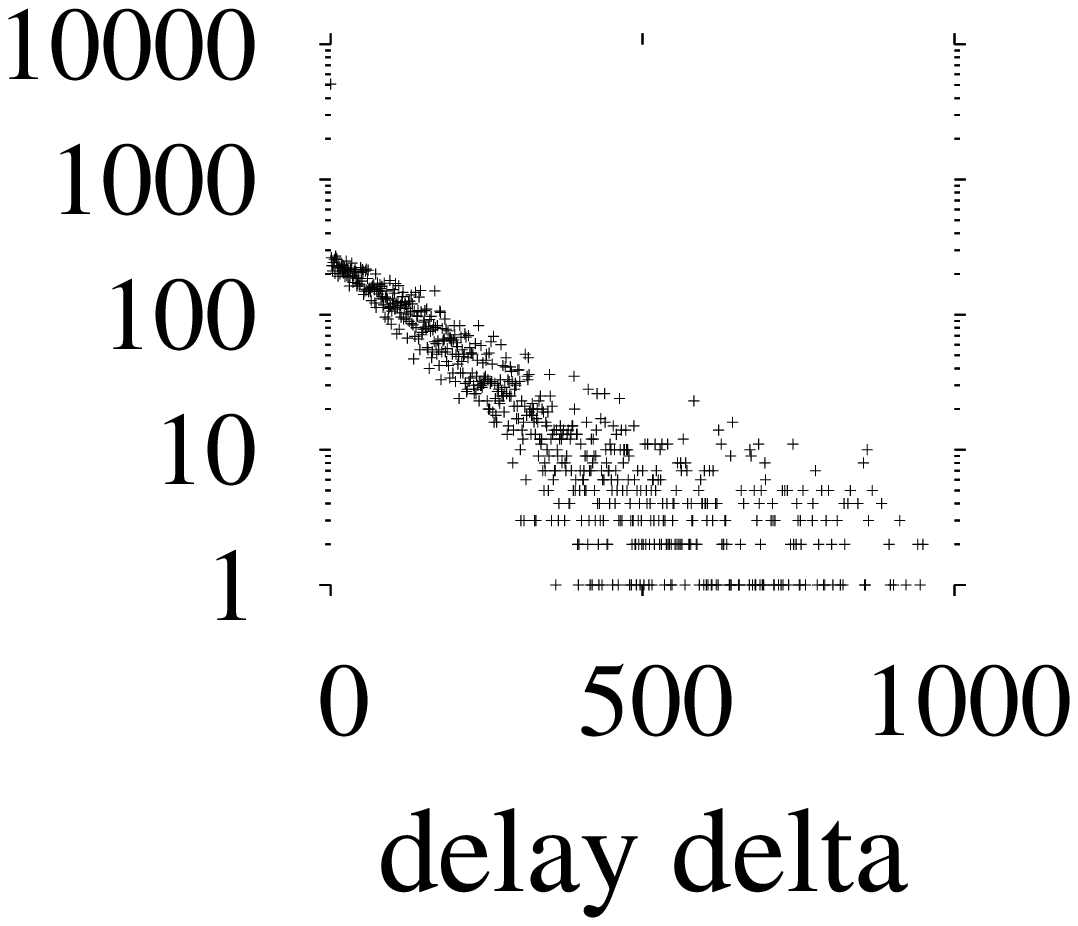}
\includegraphics[width=2.6cm]{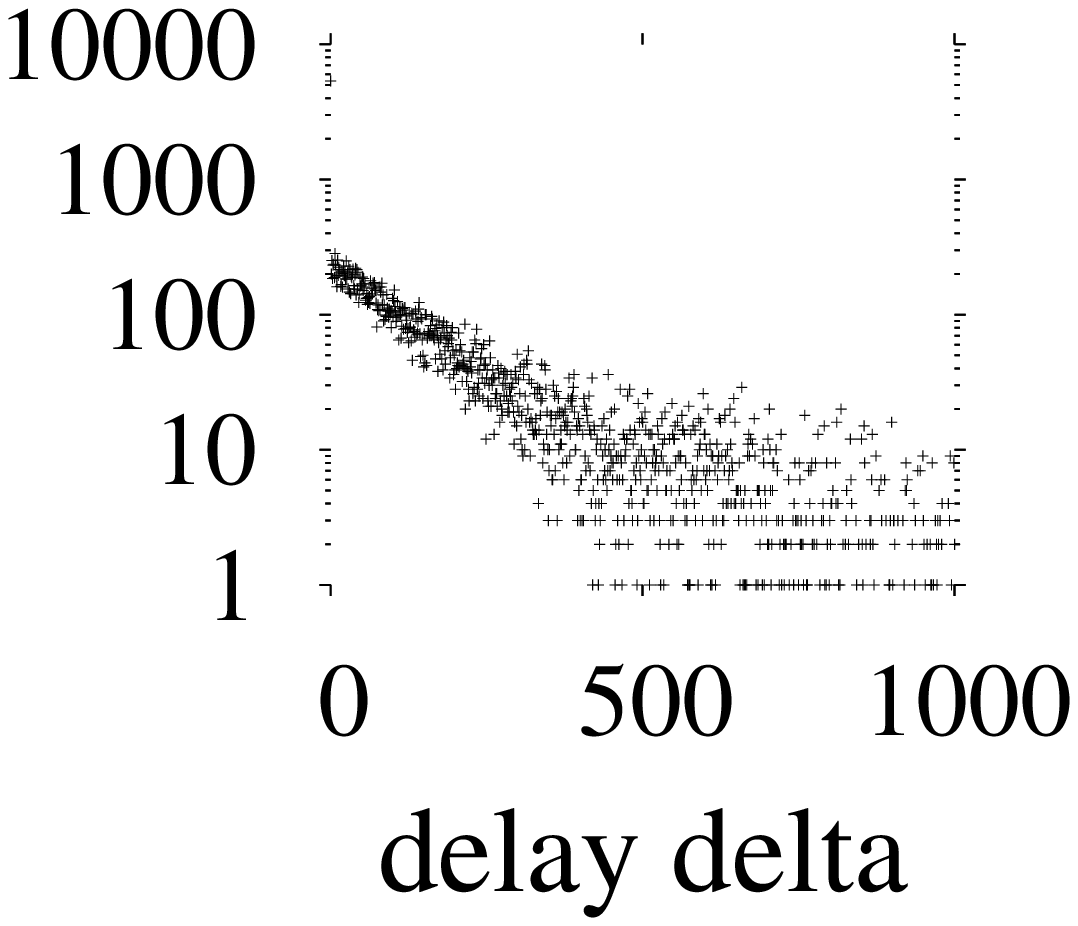}
\includegraphics[width=2.6cm]{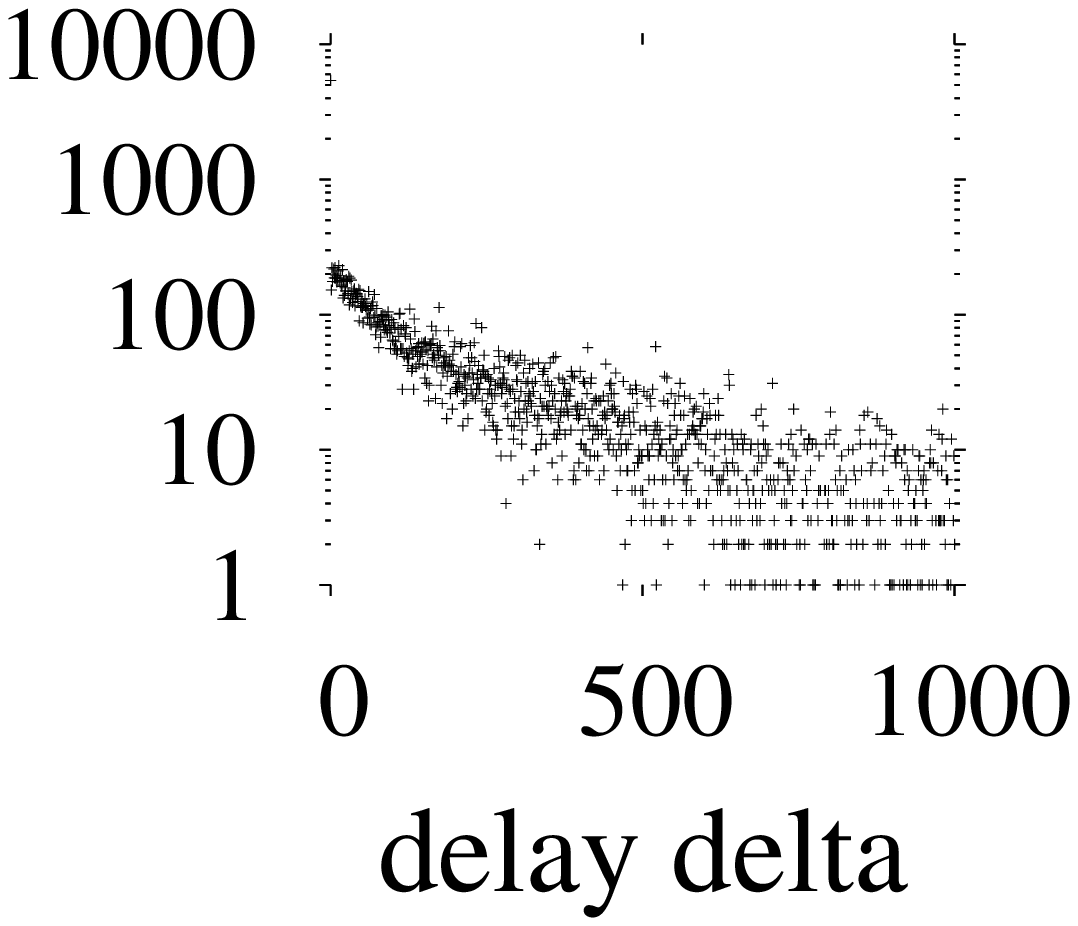}
\end{minipage}
\begin{minipage}[t]{8cm}
\vspace{-0.3cm}
\center{\textbf{Matching}\\
\vspace{-0.3cm}
\rule{8cm}{0.01cm}}
\vspace{0.1cm}
\end{minipage}
\end{center}
\caption{\label{delay_comp_flood} Delay frequency compared to Epidemic.}
\end{table}

\begin{table}[!htbp]
\begin{center}
\footnotesize
\begin{minipage}[t]{8cm}
\hspace{1cm} \textbf{$d = 1.1 $} \hspace{1.7cm} \textbf{$d = 1.5 $} \hspace{1.8cm} \textbf{$d = 2 $} \hspace{2cm}\\
\vspace{-0.8cm}
\center{\rule{8cm}{0.01cm}}
\vspace{0.1cm}
\end{minipage}
\begin{minipage}[t]{8cm}
\includegraphics[width=2.6cm]{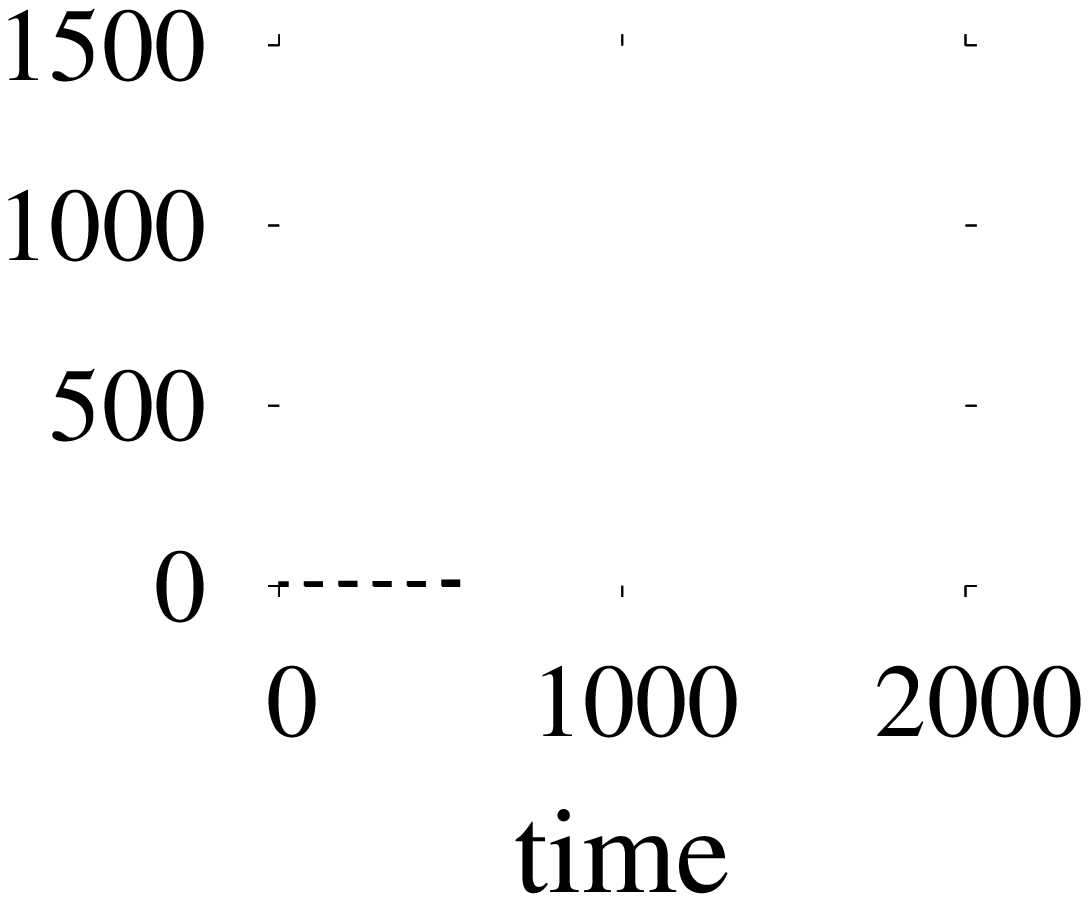}
\includegraphics[width=2.6cm]{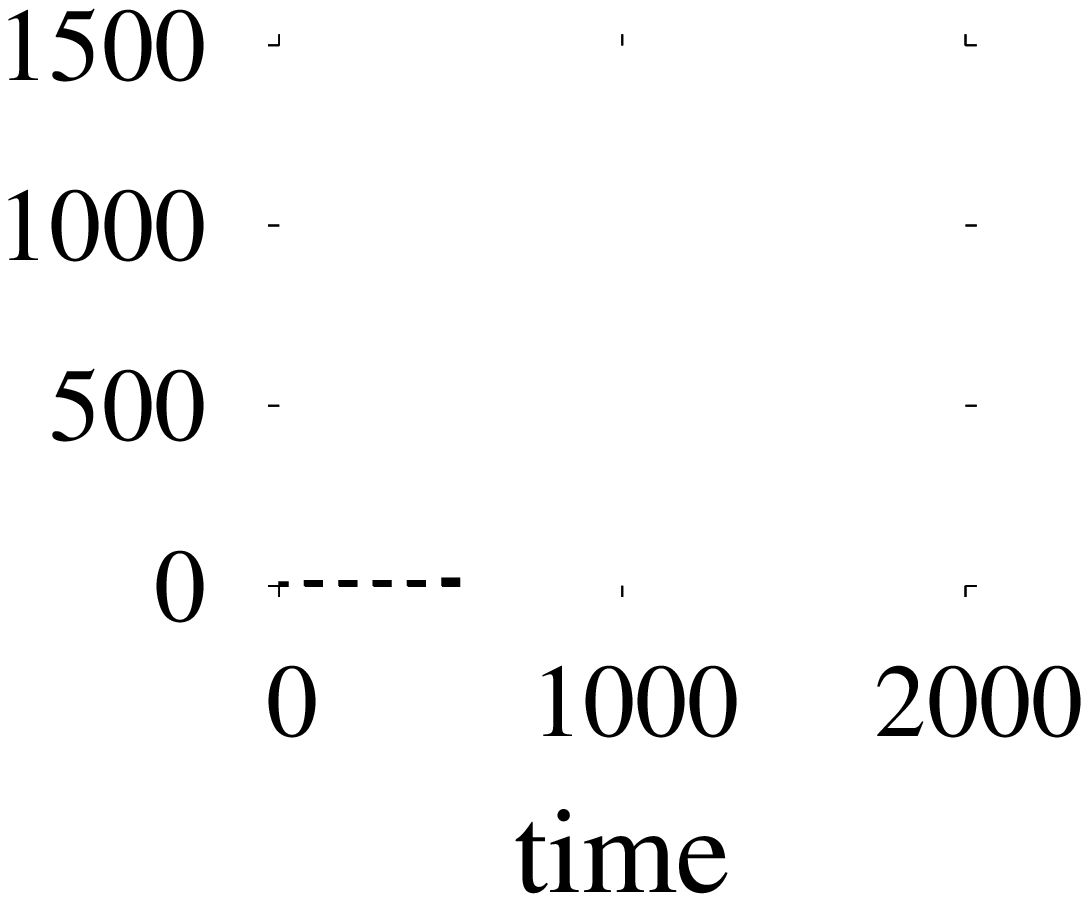}
\includegraphics[width=2.6cm]{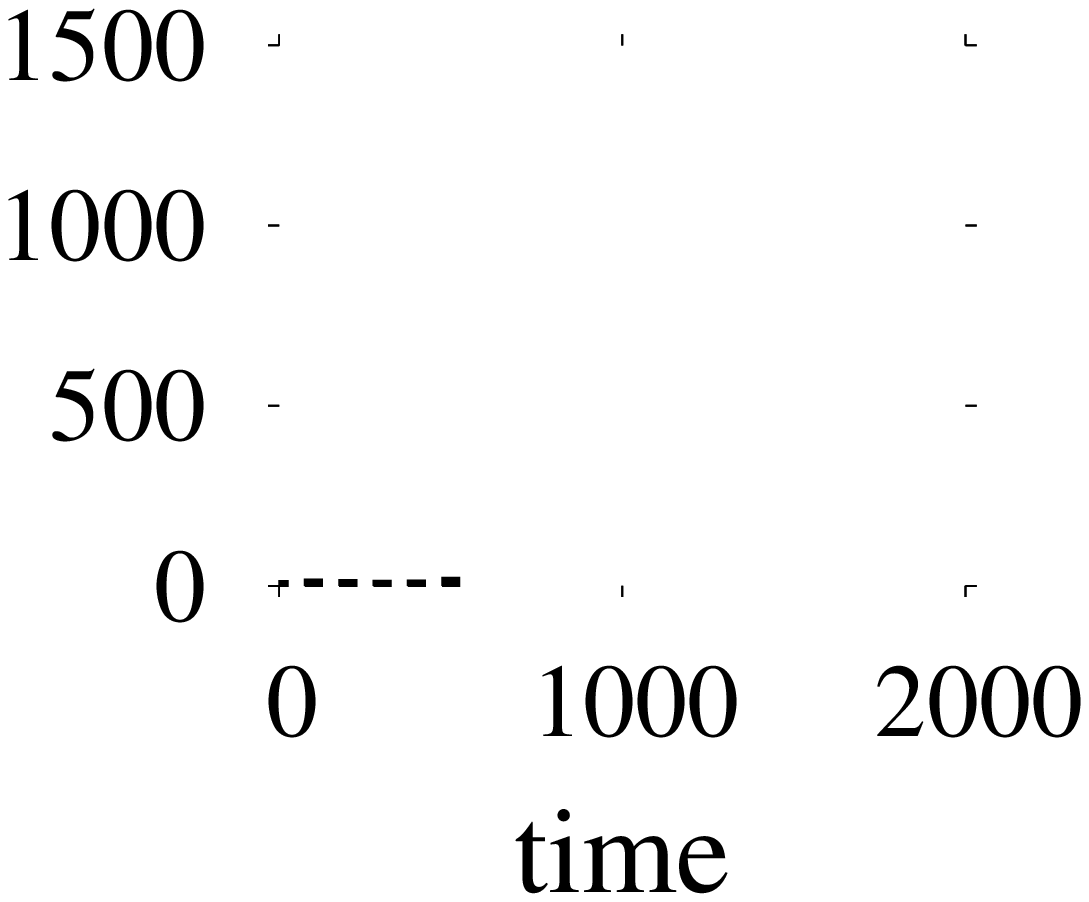}
\end{minipage}
\begin{minipage}[t]{8cm}
\vspace{-0.3cm}
\center{\textbf{Epidemic}\\
\vspace{-0.2cm}
\rule{8cm}{0.01cm}}
\vspace{0.1cm}
\end{minipage}
\begin{minipage}[t]{8cm}
\includegraphics[width=2.6cm]{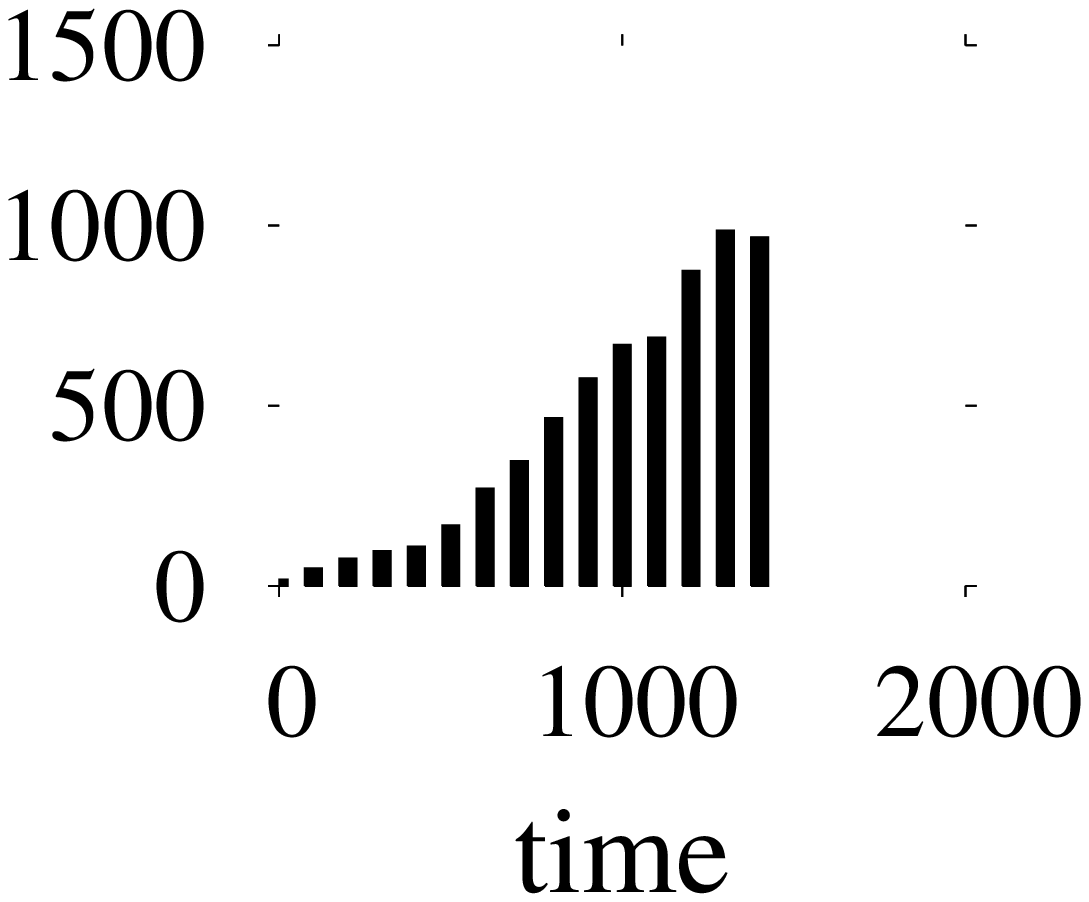}
\includegraphics[width=2.6cm]{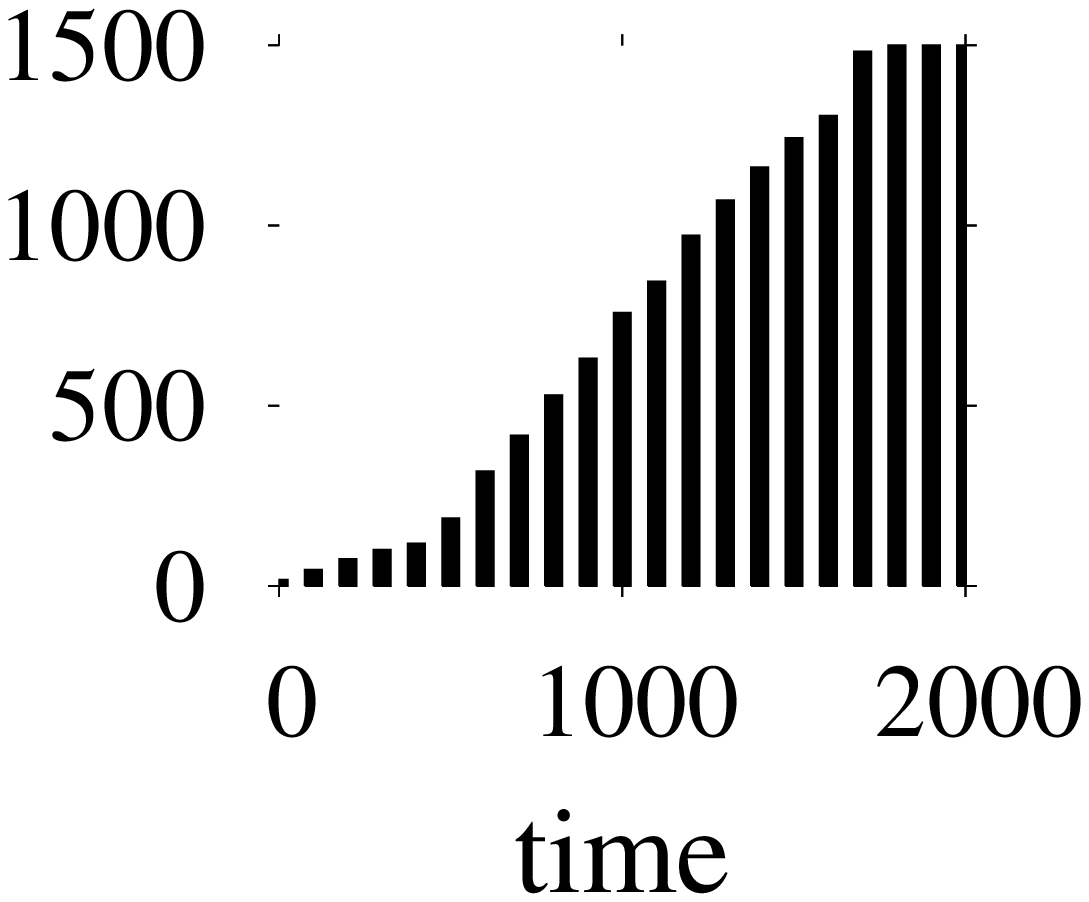}
\includegraphics[width=2.6cm]{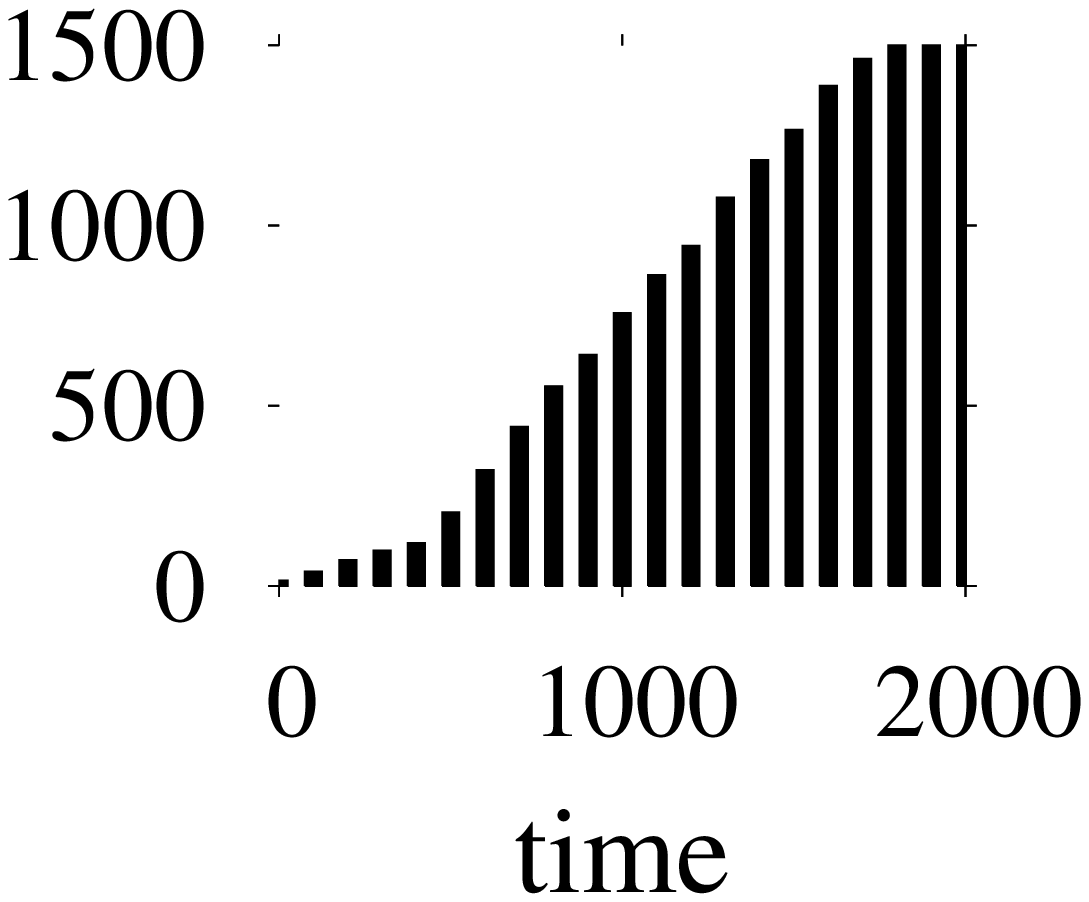}
\end{minipage}
\begin{minipage}[t]{8cm}
\vspace{-0.3cm}
\center{\textbf{Opportunistic}\\
\vspace{-0.2cm}
\rule{8cm}{0.01cm}}
\vspace{0.1cm}
\end{minipage}
\begin{minipage}[t]{8cm}
\includegraphics[width=2.6cm]{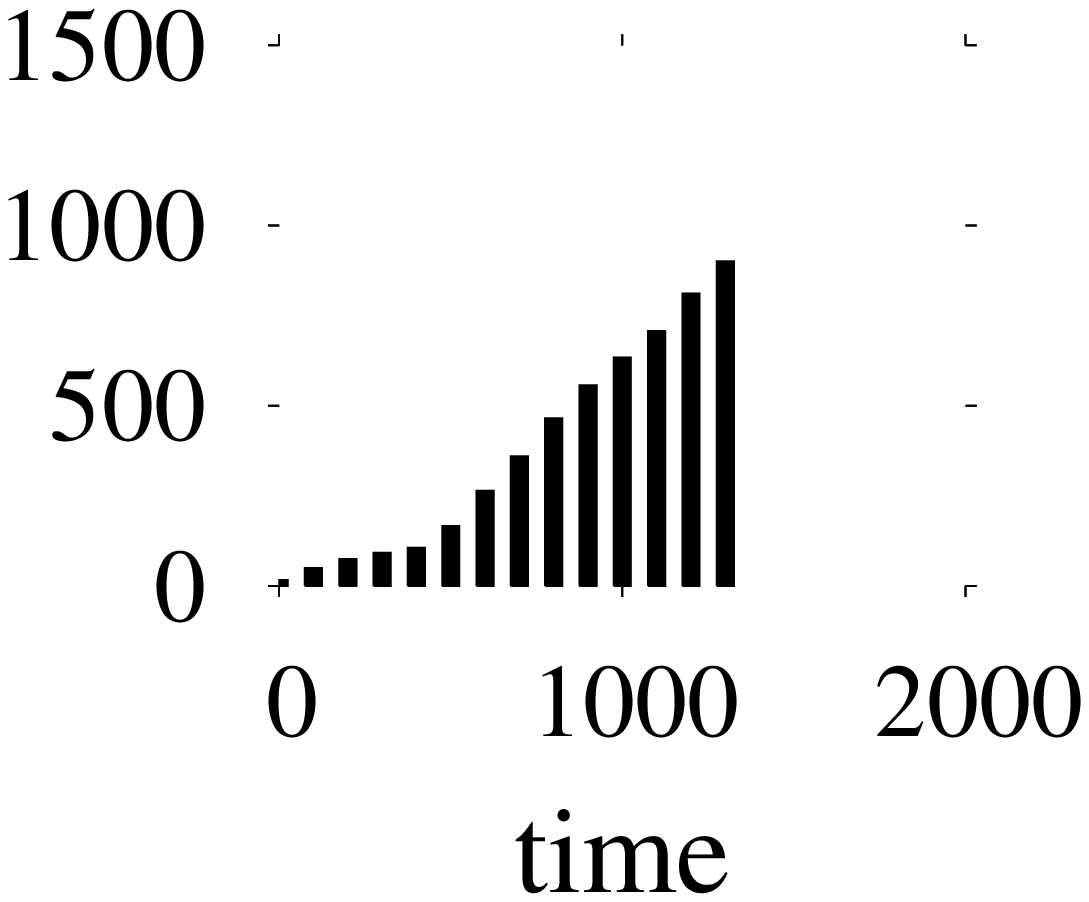}
\includegraphics[width=2.6cm]{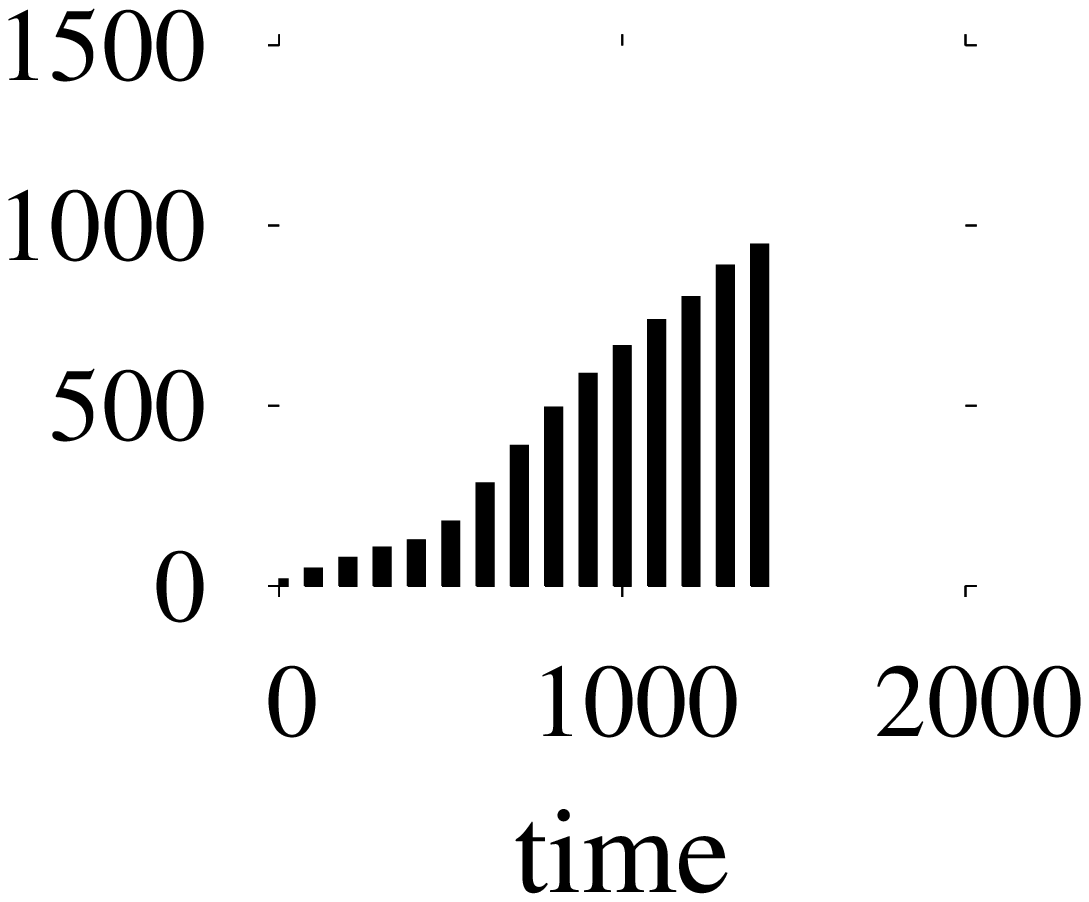}
\includegraphics[width=2.6cm]{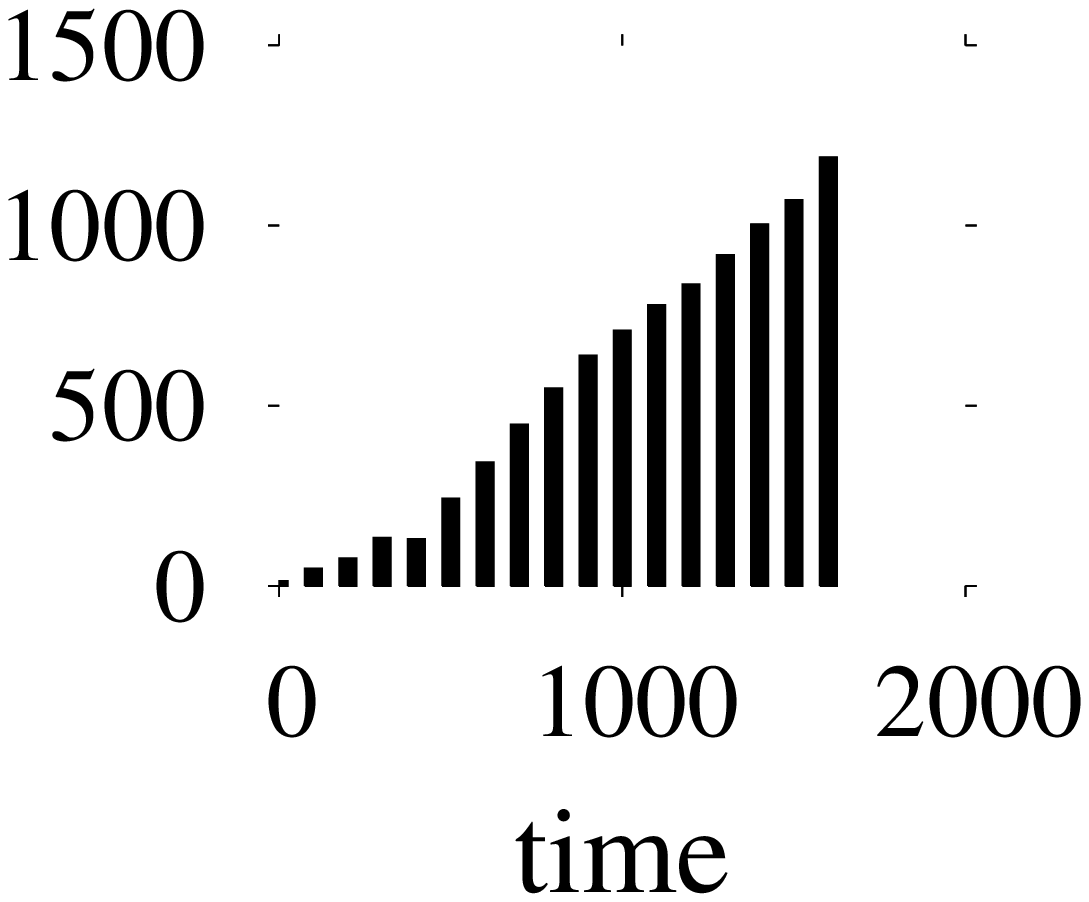}
\end{minipage}
\begin{minipage}[t]{8cm}
\vspace{-0.3cm}
\center{\textbf{Random}\\
\vspace{-0.2cm}
\rule{8cm}{0.01cm}}
\vspace{0.1cm}
\end{minipage}
\begin{minipage}[t]{8cm}
\includegraphics[width=2.6cm]{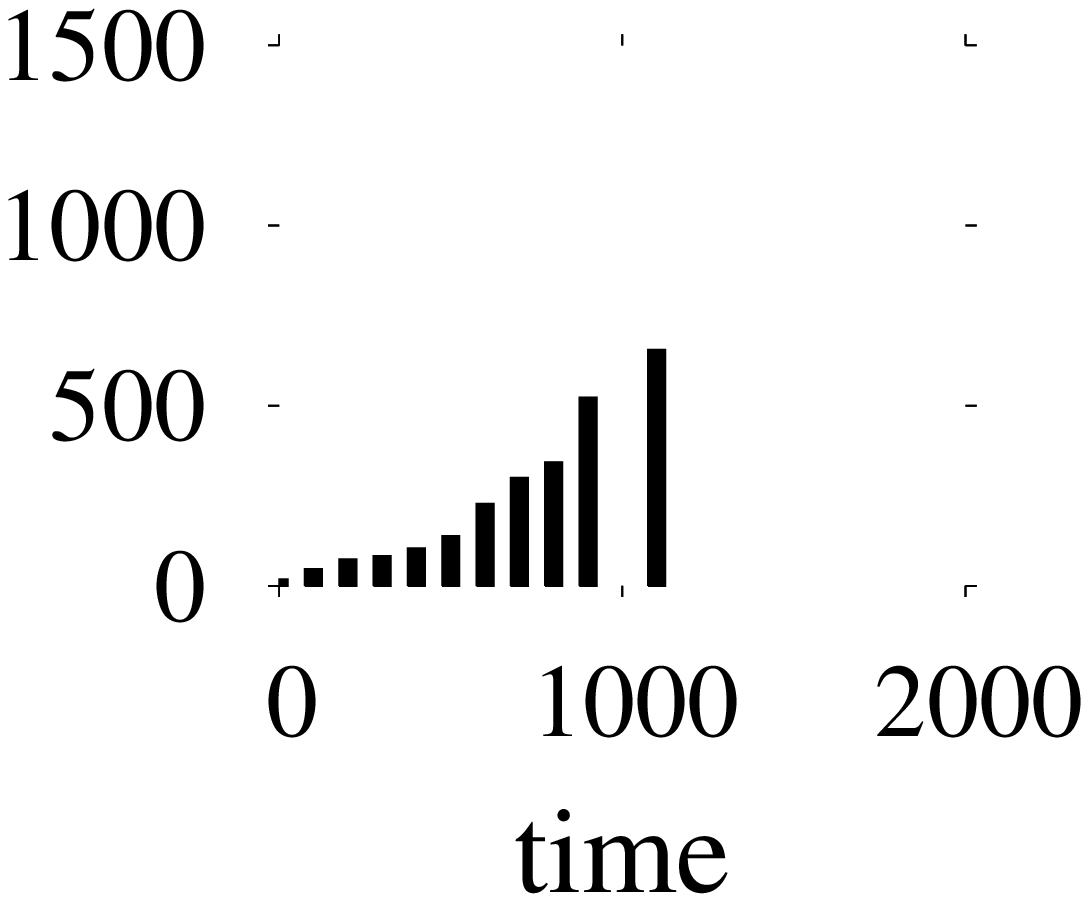}
\includegraphics[width=2.6cm]{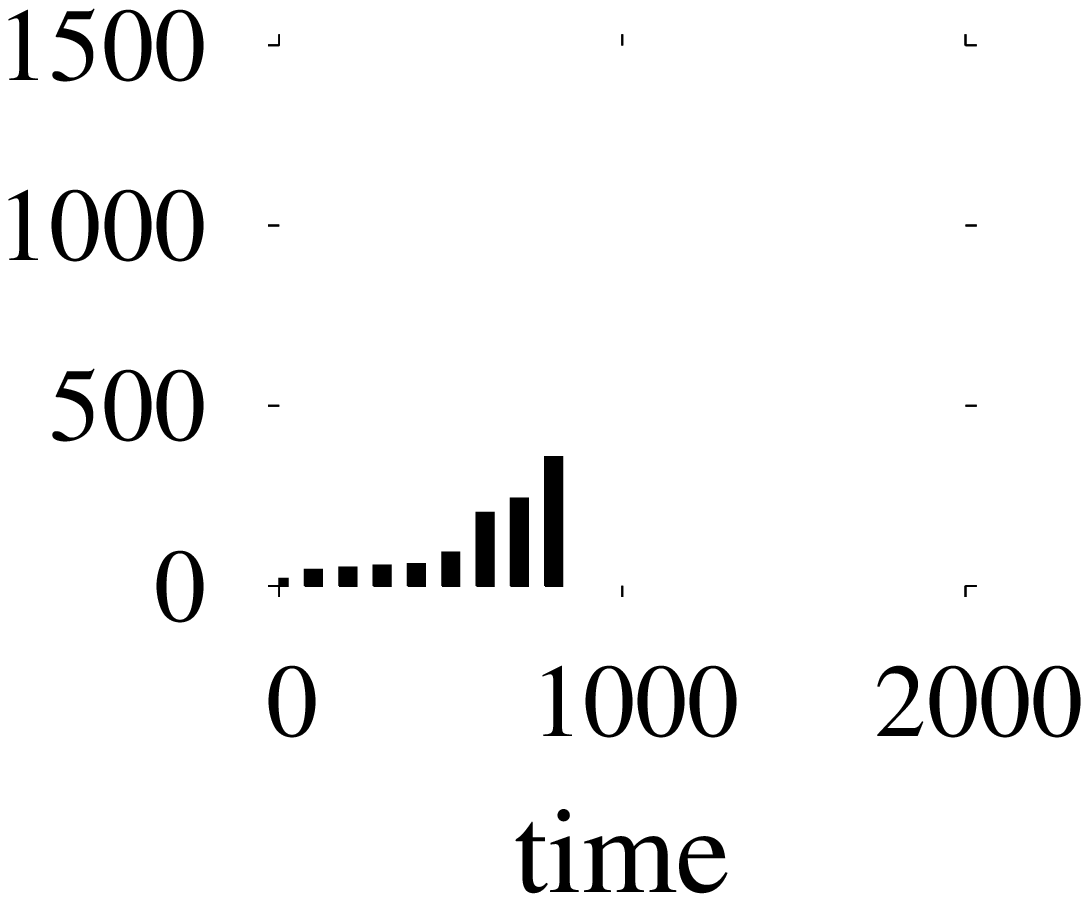}
\includegraphics[width=2.6cm]{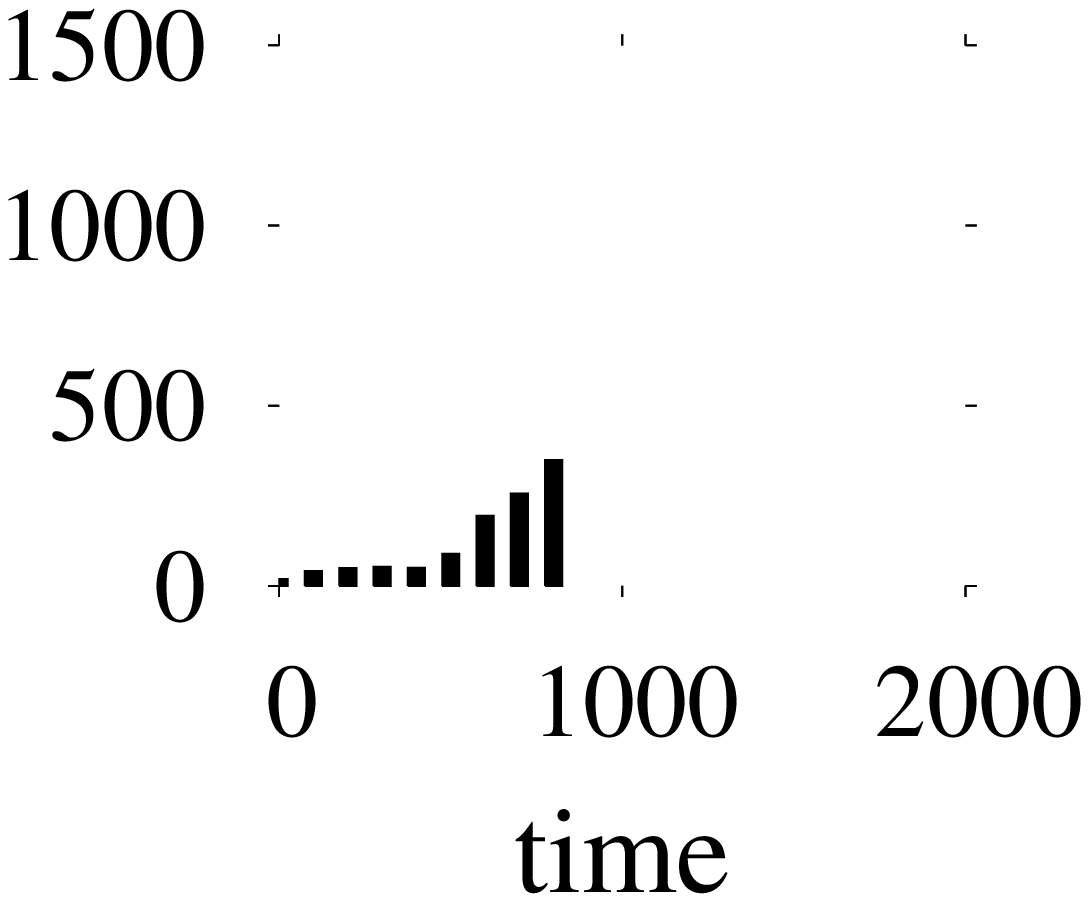}
\end{minipage}
\begin{minipage}[t]{8cm}
\vspace{-0.3cm}
\center{\textbf{Euclidean \& Angle}\\
\vspace{-0.2cm}
\rule{8cm}{0.01cm}}
\vspace{0.1cm}
\end{minipage}
\begin{minipage}[t]{8cm}
\includegraphics[width=2.6cm]{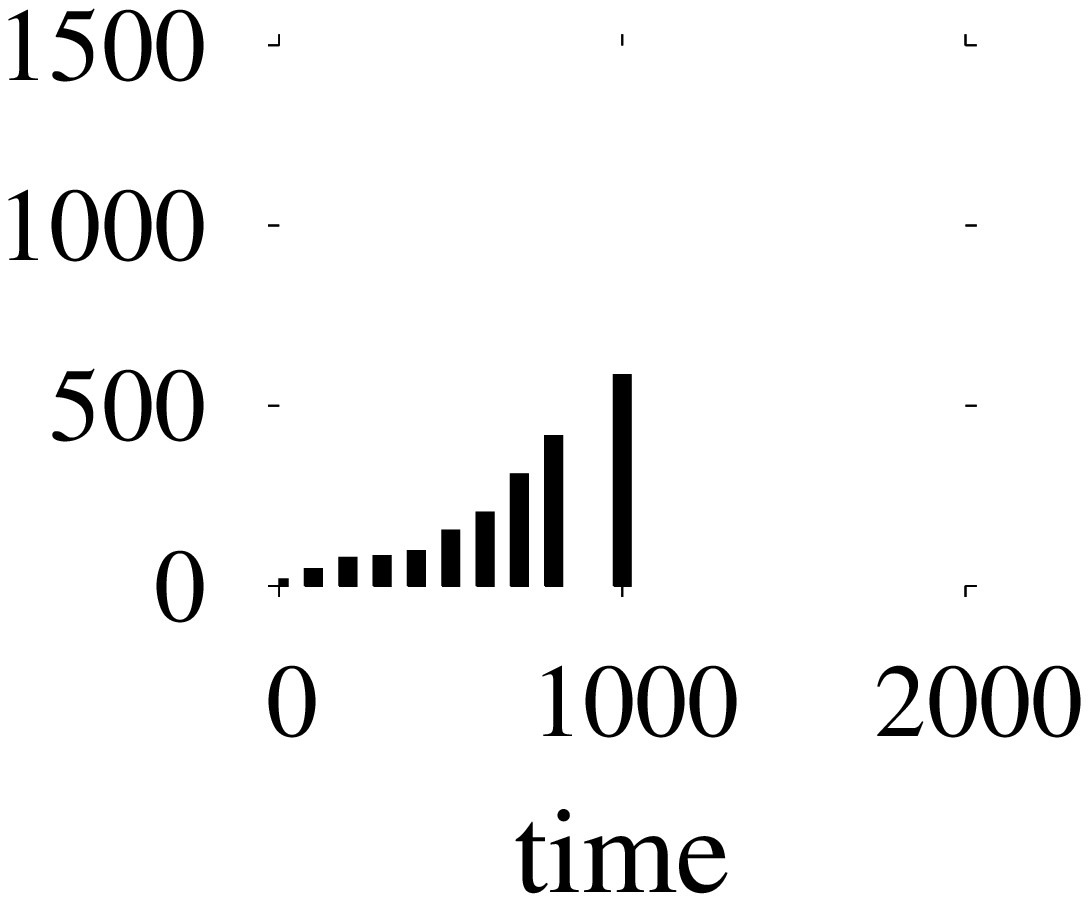}
\includegraphics[width=2.6cm]{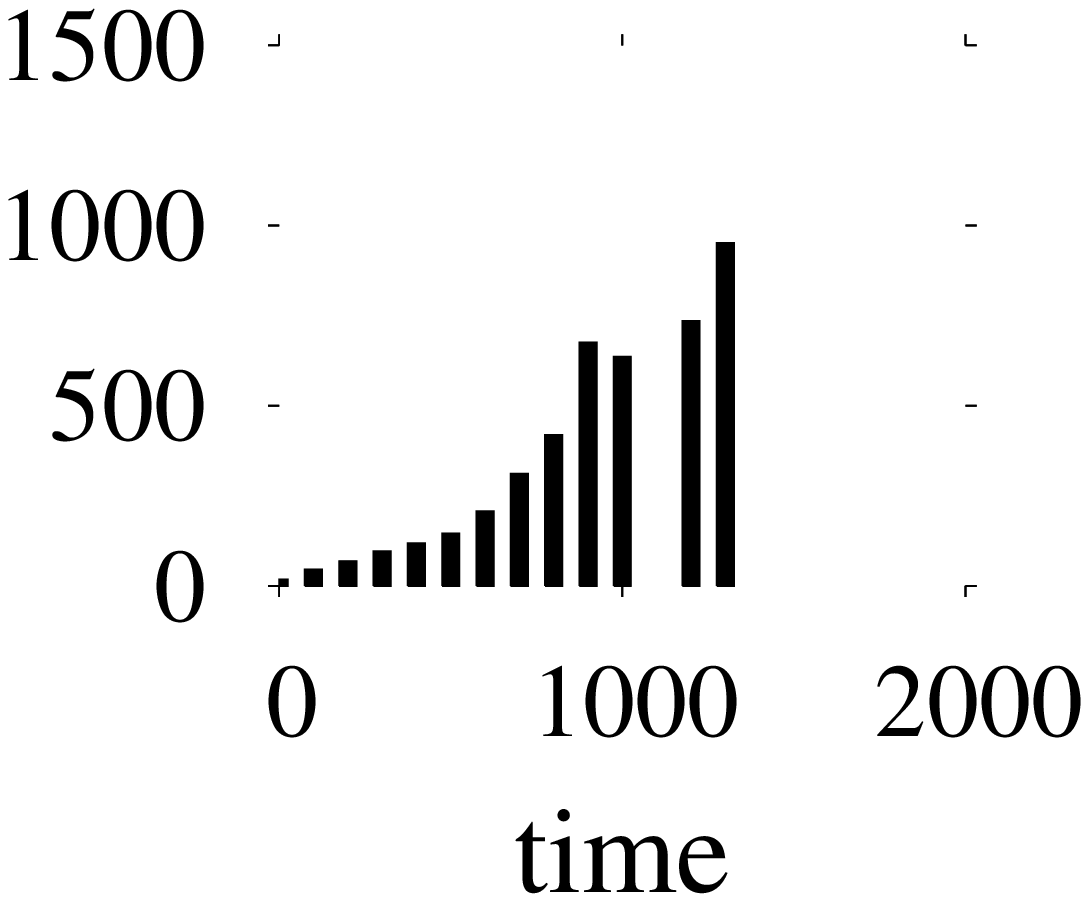}
\includegraphics[width=2.6cm]{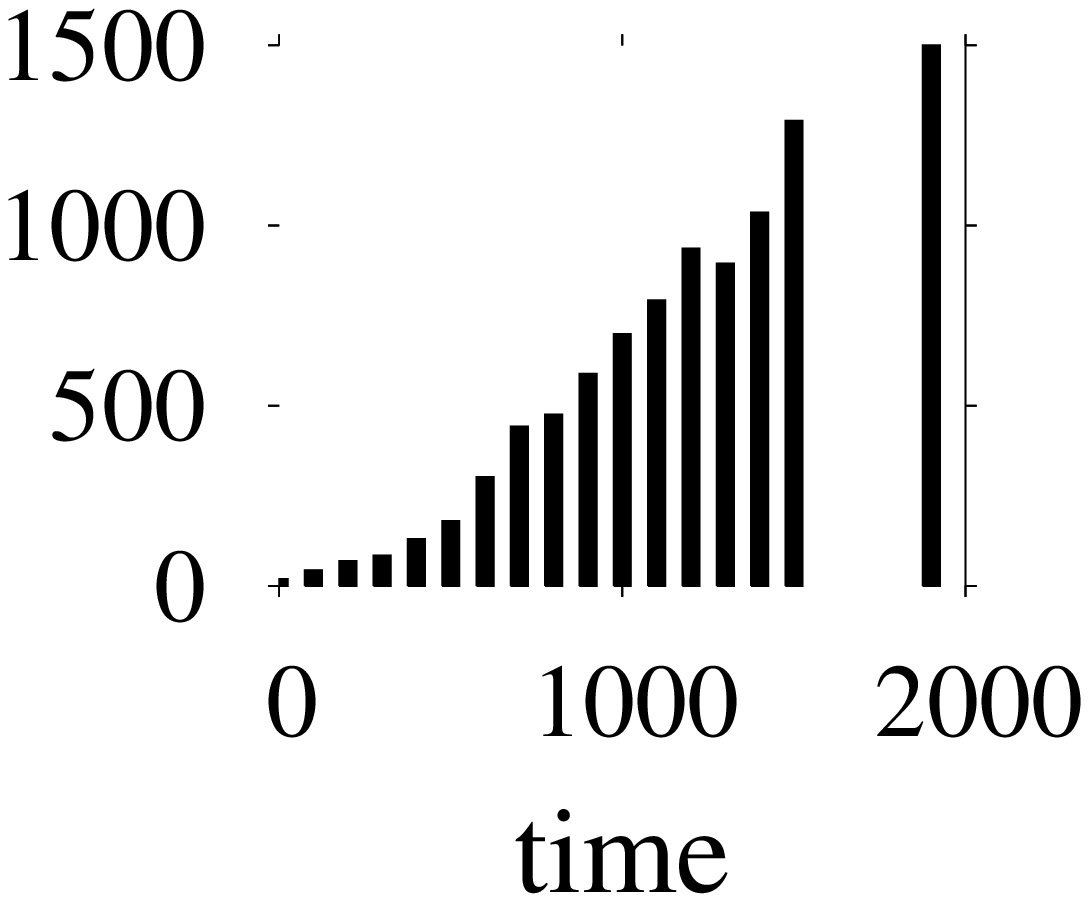}
\end{minipage}
\begin{minipage}[t]{8cm}
\vspace{-0.3cm}
\center{\textbf{Canberra}\\
\vspace{-0.2cm}
\rule{8cm}{0.01cm}}
\vspace{0.1cm}
\end{minipage}
\begin{minipage}[t]{8cm}
\includegraphics[width=2.6cm]{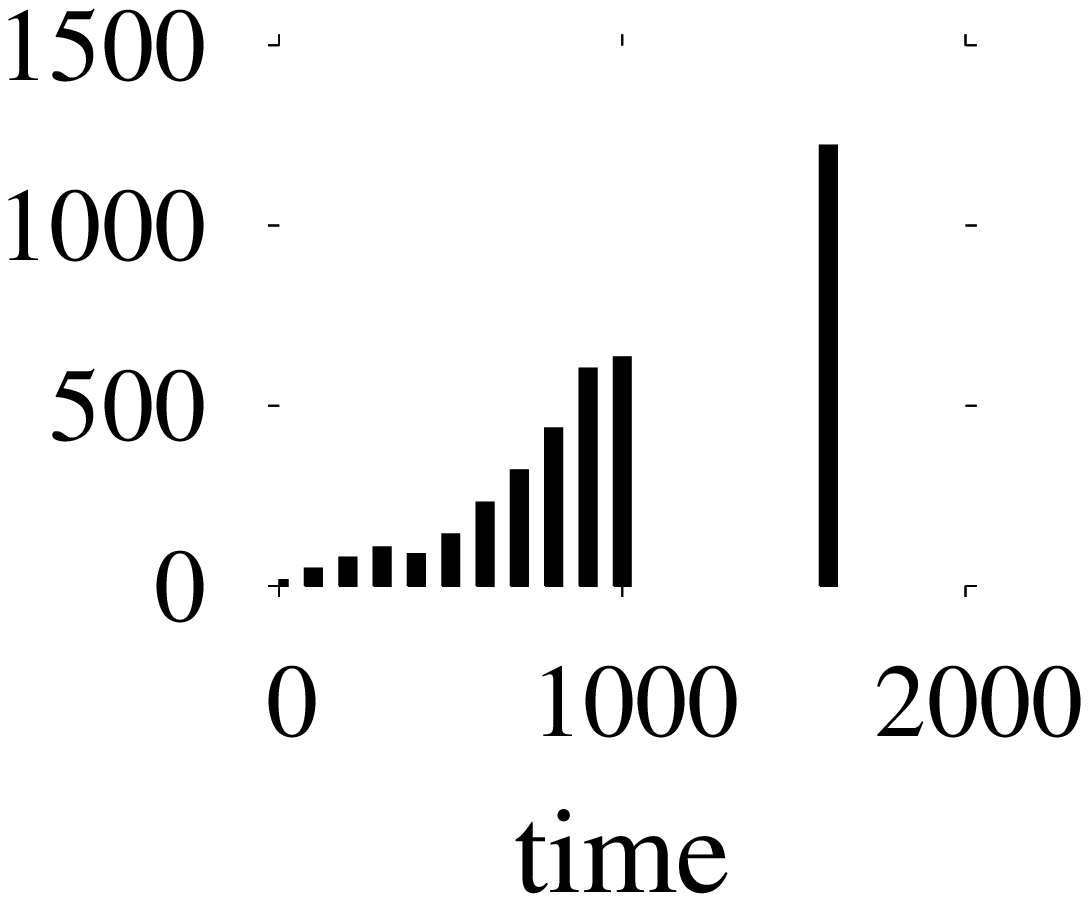}
\includegraphics[width=2.6cm]{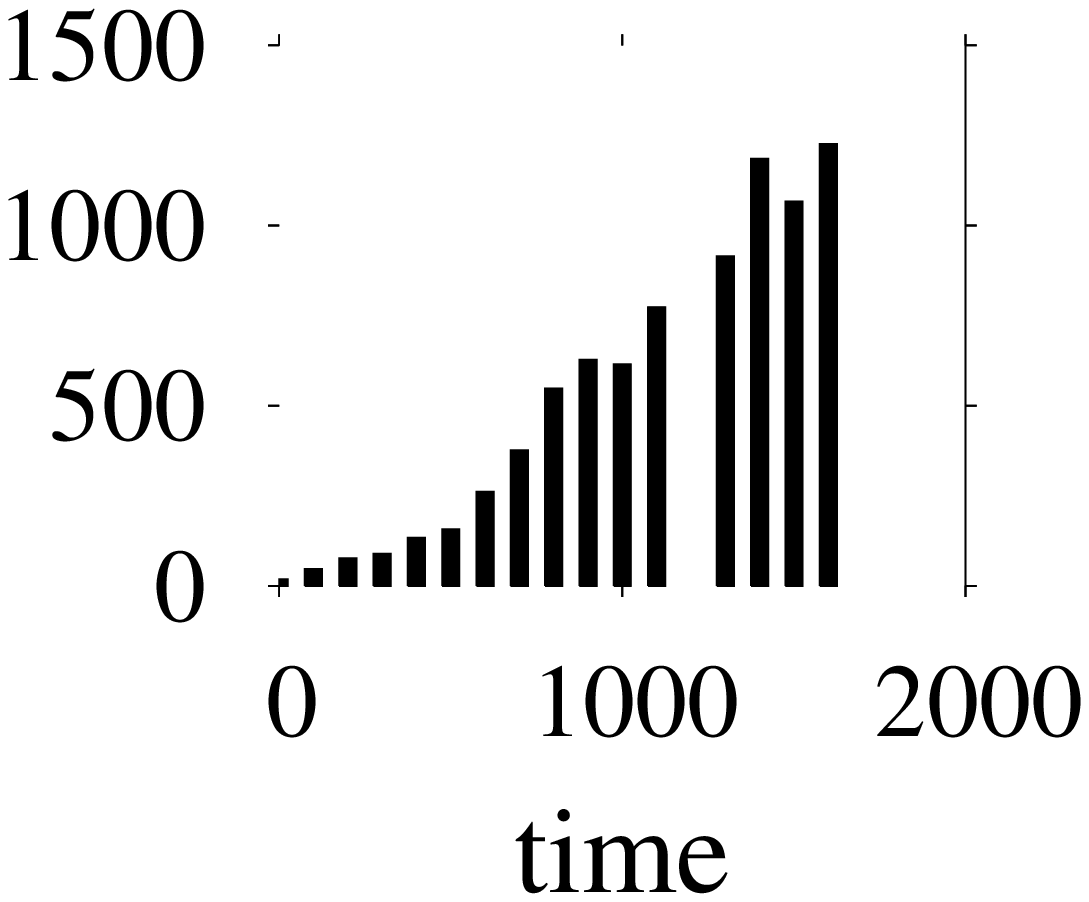}
\includegraphics[width=2.6cm]{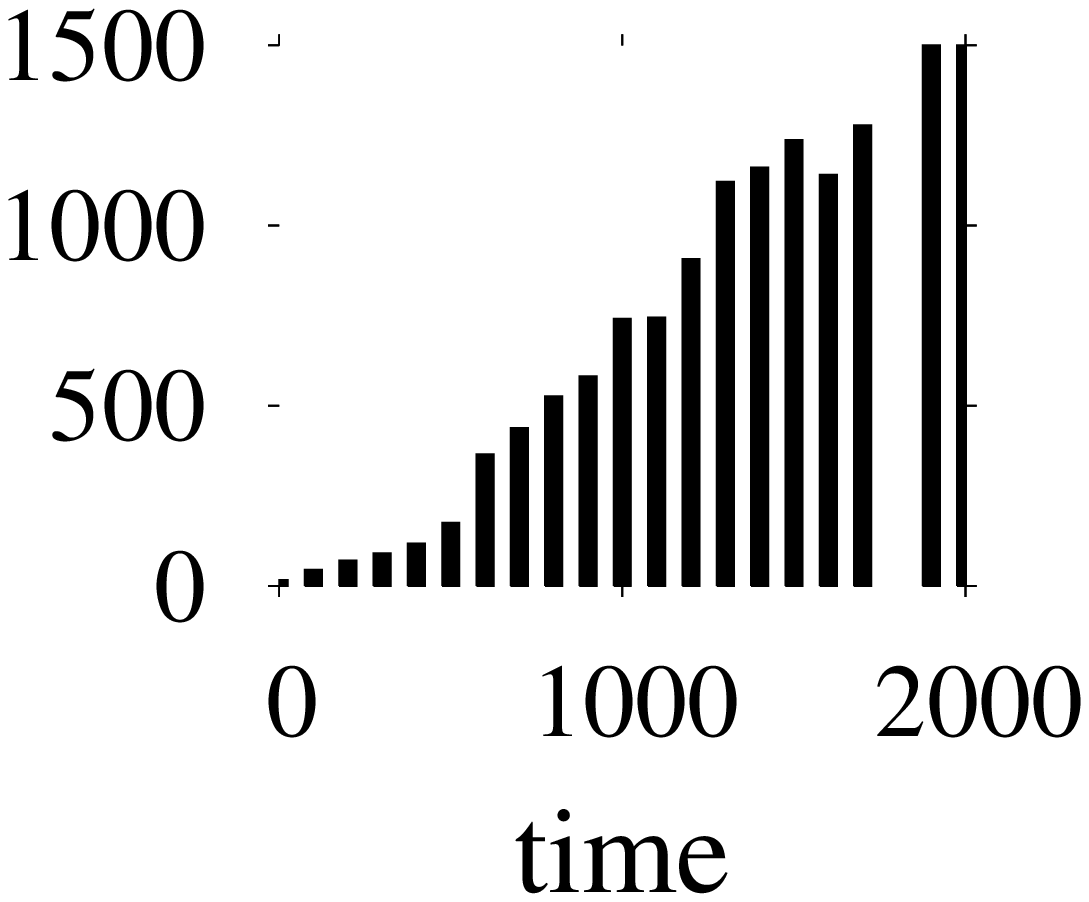}
\end{minipage}
\begin{minipage}[t]{8cm}
\vspace{-0.3cm}
\center{\textbf{Matching}\\
\vspace{-0.2cm}
\rule{8cm}{0.01cm}}
\vspace{0.1cm}
\end{minipage}
\end{center}
\caption{\label{delay_evo} Average delay evolution by steps of 100s.}
\end{table}

Table~\ref{delay_comp_flood} and Table~\ref{delay_evo} are
representative plots from one of the five experiments conducted. Note
that bins of size $1s$ have been used to aggregate data and that
$38378$ bundles have been generated in this run.
Table~\ref{delay_comp_flood} presents the frequency of the delay
difference in seconds compared to Epidemic.  For each bundle, we have
compared the delay we obtained with Epidemic to the delay we obtained
with the other algorithms. This was feasible since the traffic
generation process can be identically reproduced in our simulator with
different routing protocols.  Since it is not possible in our case to
find an algorithm having the same performances as Epidemic, we wish to
for an algorithm that approximates its performance to the extent
possible.  We can observe that the higher $d$ is, the heavier is the
tail of the distribution, except for Euclidean and Angle, which have a
much better behavior.

Table~\ref{delay_evo} shows the evolution of the average delay in
seconds over time, by steps of $100$s, for the different algorithms.
Epidemic performs the best in terms of delay and route length. It
leads to very low delays that remain constant over time. No bundles
are delivered after 500s because we stop generating traffic at that
time. For the other algorithms, bundle delay increases slightly before
500s and then linearly with time after that point because remaining
bundles are waiting in buffers to be delivered. When some bundles are
difficult to deliver, it results in delivery in batches, such as for
Matching at time 1600s for the case $d=1.1$.  Euclidean and Angle seem
to be the best solutions, resulting in low delays and fast deliveries
when the mobility patterns have power-law distributions with $d$ greater than $1.5$.\\

\subsubsection{With partial knowledge}\label{partial}

Here, we analyze the performance and the limitations of a reduction in
communication overhead that consists of nodes only diffusing the main
components of their mobility patterns.

We ran simulations for the different similarity metrics with values of
$d$ going from $1.1$ to $2$, and by taking into account only the
principal \nth{1}, \nth{2}, \nth{3}, or \nth{4} components of node's
mobility patterns.

\begin{table}[!h]
\begin{center}
\scriptsize
\begin{tabular}{|c|c|c|c|c|}
\hline
metric & & $d=1.1$ & $d=1.5$ & $d=2$\\
\hline
\textbf{Euclidean} & $l=25$ & \co{103.0}{7.7} & \co{59.1}{2.7} &  \co{54.6}{2.0} \\
& $l=4$ & \co{106.2}{4.3} & \co{60.0}{2.4} & \co{54.5}{2.3} \\
& $l=3$ & \co{107.2}{7.0} & \co{60.0}{2.4} & \co{54.9}{1.8} \\
& $l=2$ & \co{107.2}{7.0} & \co{62.2}{2.3} & \co{57.2}{1.9} \\
& $l=1$ & \co{110.7}{4.6} & \co{69.2}{3.0} & \co{75.1}{9.4} \\
\hline
\textbf{Angle} & $l=25$ & \co{103.0}{7.7} & \co{59.1}{2.7} & \co{54.6}{2.0} \\
& $l=4$ & \co{106.2}{4.3} & \co{60.0}{2.4} & \co{54.5}{2.3} \\
& $l=3$ & \co{107.4}{7.2} & \co{60.0}{2.4} & \co{54.5}{1.8} \\
& $l=2$ & \co{107.4}{7.2} & \co{62.3}{2.2} & \co{57.2}{1.9} \\
& $l=1$ & \co{110.5}{4.8} & \co{69.0}{3.0} & \co{75.1}{9.0} \\
\hline
\textbf{Canberra} & $l=25$ & \co{104.8}{4.6}  & \co{113.4}{10.4}  & \co{245.0}{41.2} \\
& $l=4$ & \co{106.4}{5.2} & \co{68.6}{4.1} & \co{98.5}{8.5} \\
& $l=3$ & \co{106.4}{7.5} & \co{68.6}{4.1} & \co{80.2}{3.3} \\
& $l=2$ & \co{106.4}{7.5} & \co{67.5}{1.6} & \co{66.2}{3.2} \\
& $l=1$ & \co{109.9}{4.5} & \co{69.9}{2.8} & \co{75.2}{9.4} \\
\hline
\textbf{Matching} & $l=25$ & \co{118.5}{5.7}  & \co{189.5}{12.1}  & \co{352.9}{56.0} \\
& $l=4$ & \co{116.2}{6.4} & \co{109.0}{4.5} & \co{225.1}{16.0} \\
& $l=3$ & \co{113.3}{4.8} & \co{109.0}{4.5} & \co{168.8}{8.8} \\
& $l=2$ & \co{113.3}{4.8} & \co{103.5}{3.3} & \co{164.7}{8.2} \\
& $l=1$ & \co{112.7}{4.1} & \co{136.8}{6.3} & \co{265.6}{10.7} \\
\hline
\end{tabular}
\end{center}
\caption{\label{res_limited.delay}Average delay. With $l$ the number of components taken into account.}
\end{table}

Table~\ref{res_limited.delay} shows the average bundle delays and
Table~\ref{res_limited.length} the average route lengths. The case
with $l=25$ in the tables comes from the results obtained with full
knowledge.  We have added here results for Angle because they are not
exactly the same as for Euclidean, even if they remain similar. We can
see two kind of behaviors for the metrics.  On one hand, for Angle and
Euclidean, the less information is available, the higher is the delay.
Route lengths remain the same, except for $l=1$ where it decreases a
bit. On the other hand, for Canberra and Matching the less information
we have, the lower are the delays (except for $l=1$ where it increases
a bit) and the lower are the route lengths. These results remain
higher in term of delay than when nodes have full knowledge of
destination's mobility patterns but show that we can obtain a
diminution of route lengths on average. This is especially the case
for Matching where route lengths are lower (up to 1 hop less) without
leading to dramatic delays for small values of $d$.

The fact that Canberra and Matching performs better in most cases with
low values of $l$ than for $l=25$ tends to confirm that these metrics
were certainly not used at their best in this study. Further analysis
of these metrics is required.

\begin{table}[!h]
\begin{center}
\scriptsize
\begin{tabular}{|c|c|c|c|c|}
\hline
metric & & $d=1.1$ & $d=1.5$ & $d=2$ \\
\hline
\textbf{Euclidean} & $l=25$  & \co{3.3}{0.0} & \co{3.2}{0.0} & \co{3.2}{0.0} \\
& $l=4$ & \co{3.3}{0.0} & \co{3.2}{0.0} & \co{3.2}{0.0}  \\
& $l=3$ & \co{3.3}{0.0} & \co{3.2}{0.0} & \co{3.2}{0.0}  \\
& $l=2$ & \co{3.3}{0.0} & \co{3.2}{0.0} & \co{3.2}{0.1}  \\
& $l=1$ & \co{3.1}{0.0} & \co{3.1}{0.0} & \co{3.0}{0.0}  \\
\hline
\textbf{Angle} & $l=25$  & \co{3.3}{0.0} & \co{3.2}{0.0} & \co{3.2}{0.0} \\
& $l=4$ & \co{3.3}{0.0} & \co{3.2}{0.0} & \co{3.2}{0.0}  \\
& $l=3$ & \co{3.3}{0.0} & \co{3.2}{0.0} & \co{3.2}{0.0}  \\
& $l=2$ & \co{3.3}{0.0} & \co{3.2}{0.0} & \co{3.2}{0.1}  \\
& $l=1$ & \co{3.1}{0.0} & \co{3.1}{0.0} & \co{3.0}{0.0}  \\
\hline
\textbf{Canberra} & $l=25$  & \co{3.3}{0.0} & \co{3.2}{0.0} & \co{3.2}{0.0} \\
& $l=4$ & \co{3.3}{0.0} & \co{3.2}{0.0} & \co{3.2}{0.0}  \\
& $l=3$ & \co{3.3}{0.0} & \co{3.2}{0.0} & \co{3.2}{0.0}  \\
& $l=2$ & \co{3.3}{0.0} & \co{3.2}{0.0} & \co{3.2}{0.0}  \\
& $l=1$ & \co{2.9}{0.0} & \co{2.9}{0.0} & \co{3.0}{0.0}  \\
\hline
\textbf{Matching} & $l=25$  & \co{2.5}{0.0} & \co{2.5}{0.0} & \co{2.4}{0.0} \\
& $l=4$ & \co{1.9}{0.0} & \co{1.8}{0.0} & \co{1.9}{0.0}  \\
& $l=3$ & \co{1.7}{0.0} & \co{1.8}{0.0} & \co{1.8}{0.0}  \\
& $l=2$ & \co{1.7}{0.0} & \co{1.7}{0.0} & \co{1.7}{0.0}  \\
& $l=1$ & \co{1.5}{0.0} & \co{1.5}{0.0} & \co{1.5}{0.0}  \\
\hline
\end{tabular}
\end{center}
\caption{\label{res_limited.length}Average route lengths. With $l$ the number of components taken into account.}
\end{table}

\section{Conclusion and future work}\label{sec_conclu}

The main contribution of this paper has been the definition of a
generic routing scheme using the formalism of a high-dimensional
Euclidean space constructed upon mobility patterns.  We have shown
through a scenario inspired from reality that it can applied to DTNs
and that it can bring benefits in terms of bundle delay and
communication costs. The idea was to apply a well know and powerful
artifact for routing in DTNs. We claim through this study to have
opened new perspectives for routing.  Much work remains to be done.

Future work along these lines might include studies concerning the
impact of the structure of the Euclidean space, i.e., the number and
type of dimensions, and the similarity function.  Different kind of
Euclidean space have to be investigated by considering schemes like
the one describes in Sec.~\ref{concept} that takes for each dimension
the frequency of contacts between a certain pair of nodes.  It might
also include studies around other parameters like the number of nodes
and their type of mobility patterns because it may play a great role
in the performances of the routing scheme. What happens for instance
if because of dynamics in mobility patterns, these patterns are
constantly evolving and shifting? Furthermore we have consider in this
work that nodes have the knowledge of their mobility pattern, but to
fit better with real conditions of deployments questions about the way
nodes learn or get their mobility patterns have to be investigate.

Then, the Euclidean spaces that we have studied here are finite in terms of number of
dimensions, but to be implementable in really in some cases these spaces may have 
to be consider as infinite. For instance, in the space we use as a case study in Sec.~\ref{case_study}, 
the number of locations can rise up to the infinite. It can be feasible if nodes are able 
to understand each other when they exchange information about their mobility pattern. Thus, 
additional semantics features have to be found to allow this.

Finally, in this work we consider that the essential characteristics of a node's mobility or
contact patterns are fully captured by the frequency with which
nodes find themselves in certain locations or the probability that
they will be in proximity to certain nodes.  However, prior work~\cite{routingdtn}, 
has demonstrated the interest of capturing temporal
information as well.  It is well known that network usage patterns
follow diurnal and weekly cycles.  We could easily imagine two nodes
that visit the same locations with the same frequencies, but on
different days of the week.  Though it still might make sense to
route to one node in order to reach the other, especially if there
is a relay node at the commonly visited location, the Euclidean
space description that we have provided does not capture this point
of dissimilarity.  We can, however, imagine ways in which the
dimensional representation could capture temporal information as
well.  For instance, visiting patterns could be translated into the
frequential domain.  A node's visits to a location could be
represented by a point on a frequency axis, capturing the dominant
frequency of visits, and a point on a phase axis, as well as a point
on the axis already described, that represents the overall
probability of visiting the location.

\bibliographystyle{IEEE}

\end{document}